\documentclass[12pt]{article}
\pdfoutput=1
\usepackage[utf8]{inputenc}
\usepackage[english]{babel}
\usepackage[sort&compress,square,comma,numbers]{natbib}
\usepackage{amsmath,amssymb,graphicx,slashed}
\usepackage{placeins}
\usepackage{cancel}
\usepackage{multirow} 
\usepackage{threeparttable}
\usepackage{nccmath}
\usepackage{mathrsfs} 
\usepackage{braket} 
\usepackage{setspace}
\usepackage{xfrac}
\usepackage{array}
\usepackage{appendix}
\usepackage{etoolbox}
\usepackage{feynmp-auto}
\usepackage{caption}
\usepackage{subcaption}                                
\usepackage{adjustbox}
\usepackage{float}
\usepackage[tableposition=below]{caption}
\usepackage{longtable}
\usepackage{color}			
\usepackage{verbatim}
\usepackage{booktabs}
\usepackage{labelschanged}
\usepackage{titlesec}
\definecolor{gray1}{gray}{0.55}
\titleformat{\chapter}[hang]{\Huge\bfseries}{\thechapter\hspace{10pt}\fontsize{100}{100}\textcolor{black}{$\diamond$}\hspace{20pt}}{0pt}{\Huge\bfseries}
\titleformat{\section}[hang]{\large\bfseries}{\thesection\hspace{6pt}\textcolor{gray1}{$\circ$}\hspace{10pt}}{0pt}{\large\bfseries}
\usepackage[
	colorlinks=true,
	citecolor=red,
	linkcolor=blue,
	urlcolor=blue,
	hypertexnames=false]{hyperref}
\newcommand{\LL}{\mathcal{L}}

\newcommand{\OO}{\mathcal{O}}

\newcommand{\PP}{\mathcal{P}}

\newcommand{\arXiv}[2]{\href{http://arxiv.org/pdf/#1}{{\tt #2/#1}}}
\newcommand{\arXivold}[1]{\href{http://arxiv.org/pdf/#1}{{\tt #1}}}
\newcommand{\beq}{\begin{eqnarray}}
\newcommand{\eeq}{\end{eqnarray}}

\begin{document}
\begin{titlepage}
\vspace{1cm}
\begin{center}
		\LARGE \bf 
		Multi-Scalar Production At Large Center-Of-Mass Energy
			
\end{center}
	\vskip .3cm
	
	\renewcommand*{\thefootnote}{\fnsymbol{footnote}}

\vspace{0.9cm}
\begin{center}
		
		\bf
		Ali Shayegan Shirazi
\end{center}
	
	\renewcommand{\thefootnote}{\arabic{footnote}}
	\setcounter{footnote}{0}


\begin{center} 

 	{\emph{Instituto de Alta Investigaci\'{o}n,\\ Universidad de Tarapac\'{a},\\ Casilla 7D, Arica, Chile}}

\end{center}

\vspace{1cm}

\centerline{\large\bf Abstract}
\begin{quote}
In quantum field theory, the probability of producing scalar particles grows factorially as a function of the number of the particles produced. This poses a problem theoretically, in maintaining unitarity, and is counter-intuitive phenomenologically. The factorial growth is a byproduct of the perturbation theory, but has been found in some of the semi-classical and non-perturbative calculations as well. Recently, it has been proposed that the factorial growth might be observable in the future 100 TeV hadron collider. After reviewing some of the past calculations, we analyze the cancelation of IR divergences to find Altarelli-Parisi function and the Sudakov form factor. We then use these to write an equation for the generating function of scalar jet rates. We further argue that we can turn the jet rates into particle cross section by swapping the opening angle with particle mass. We will present our results for $\phi^3$ theory in four and six spacetime dimensions, and $\phi^4$ in four spacetime dimensions. We find that, while the final particles are relativistic, the cross sections do not grow factorially.

 \end{quote}
\end{titlepage}

\tableofcontents
\newpage
\section{Introduction}\label{sec:introMULTI}
It has been known for a while that the production of many-particle states at high energies can become a factorial function of the final number of particles. This problem drew attention in the Electro-Weak theory when it was found that the B+L-violating cross section, via instanton-like processes, become large at high energies with large number of bosons in the final state~\cite{Ringwald:1989ee, Espinosa:1989qn}. Motivated by this discovery, the effect was analyzed in scalar theories to find whether the ever-increasing cross section is actually physical and what is the upper bound on the cross section without violating unitarity~\cite{zakharov, Argyres:1992np}. Multiple methods were used in order to find either the amplitude or the cross section with many bosons in the final state. Using generating functional method, Brown found the exact amplitude at threshold for any number of final states~\cite{Brown:1992ay}. The method were applied to the loops and summing the loop corrections to all order~\cite{Smith:1992rq, Voloshin:1994yp,Libanov:1994ug, Libanov:1997nt,Schenk:2021yea}. Other methods includes, but are not limited to, recursion relations~\cite{Argyres:1993wz, Papadopoulos:1993aw}, coherent state formalism~\cite{Son:1995wz, Khoze:2018mey,Demidov:2022ljh}, instanton calculation (Lipatov method)~\cite{Lipatov:1977hj, maggshif91, maggshif92}, and functional Shr\"odinger equation~\cite{cornwall92,Cornwall:1993rh}. 

 Most of these research are from before LHC; where we did not observe any B-L violation contrary to what we expected from EW enhancement of such cross sections. The problem had been forgotten for a decade or two. In the past few years, however, perhaps because the 100 TeV collider is getting closer to reality, the multi-particle cross section, in particular multi-Higgs cross section, has received new attention~\cite{Khoze:2014kka, Degrande:2016oan, Khoze:2017tjt, Khoze:2017ifq,Khoze:2018mey,Khoze:2015yba}. According to~\cite{Khoze:2017tjt}, the factorial growth is in fact physical and observable in experiment. Consequently, they foreseen new phenomenon in high energy colliders (of order 100 TeV or so) with many-particles signature. This phenomenon has been dubbed ``Higgsplosion". They propose that the unitarity can be preserved through a mechanism called `` Higgspersion". Moreover, it is claimed that these mechanisms can solve the Hierarchy problem. 
 
 It is debatable whether the Higgsplosion is physical and whether the calculation is conclusive~\cite{Voloshin:2017flq,Khoze:2018qhz, Monin:2018cbi,Dine:2020ybn, Belyaev:2018mtd,Curko:2019dtu}. The inclusion of the heavy fermion loops can change the amplitude at high multiplicity of particles, unitarity might not be actually preserved in this mechanism, etc. We also believe that there is a great weight given to the perturbative calculation which in our opinion should not be trusted at the point where the amplitude becomes large (thought, it is claimed that non-perturbative calculation valid for any $\lambda n$ also gives factorial growth).
 
 Most of the calculations referred above are based on the multi-scalar cross section near threshold, i.e. when the final particles are non-relativistic and are at fixed angle with respect to each other, hence free of collinear divergences. This will confine these results to a small corner of the phase-space. What we will try to do in this paper is to focus on the region of phase-space where there are enhancements in the amplitude due to collinear divergence's. That is, we want to address the phenomenology of multi-scalar by finding the cross section for producing $n$ scalar jets. This is done using Jet Generating Functional (JGF)~\cite{Dokshitzer:1991wu, Ellis:1991qj}, which satisfies an equation analogous to Dokshitzer-Gribov-Lipatov-Altarelli-Parisi (DGLAP) equation~\cite{Altarelli:1977zs, Lipatov:1974qm}. These calculations are semi-non-perturbative in nature, and as a result we expect them to be accurate in the large multiplicity limit. 

 The main object in our calculation is the Sudakov form factor~\cite{Sudakov:1954sw,collins:1989}, which is the probability that a particle would not split into two or more particles, i.e. the probability that it stays distinct, given the resolution of the experiment. We start with a particle at energy of order $t\approx Q^2$, where $Q$ is the hard scale, e.g. the center-of-mass energy, and an IR regulator $t_0$, which can be the minimum opening angle required to resolve two lines, e.g. $t_0=t \delta^2$, or, as we will discuss below, in the massive theory, if the opening angle is not too large, it is $t_0=m^2$. Given these two scales, the Sudakov factor, $\Delta$, is of the form
 \begin{equation}\nonumber
 \Delta(t,t_0)=\exp\bigg(-\int_{t_0}^tdz\PP(z,t')dt'\bigg).
 \end{equation}
where $\PP$ is the Altarelli-Parisi (AP) splitting function and $z$ is the energy fraction carried by the daughter particle. 
 
 We will find the AP function and the Sudakov factor for the cubic theory in four and six spacetime dimensions, and for the quartic theory in four dimensions. We will then use these to write a JGF for these theories and use it to find the jet rates. To do so it is insuring and instructive to check that the IR divergences cancel.

In section~\ref{sec:reviewpert}, we briefly review the perturbative and semi-classical/non-perturbative methods that had been used in the past in regards to multi-scalar cross sections. We also review the phenomenology of the Higgsplosion, where it is claimed that the multi-Higgs production become unsuppressed due to the factorial growth and observable in the future colliders. In section~\ref{sec:jetGF}, we discuss the jet generating function method and apply it to the scalar theories in sections~\ref{sec:phi36D}, \ref{sec:phi44D}, and \ref{sec:phi34D} for $\phi^3_{6D}$, $\phi^4_{4D}$, and $\phi^3_{4D}$, respectively. In section~\ref{sec:offsh}, we apply our results to the process $\phi^*(\text{off-shell})\to n\phi$. The discussion and conclusion is presented in section~\ref{sec:conclusion}.


\section{Review of the Past Calculations}\label{sec:reviewpert}

\subsection{Generating Function Method}
 Brown~\cite{Brown:1992ay} has utilized an elegant and simple method for finding the amplitude for an off-shell scalar to produce $n$ on-shell particles, when the particles are at threshold. The idea is that, when the final particles are at threshold, the generating function of the tree amplitudes is the classical solutions of the equation of motion in the presence of a source term, and using the classical solution one finds the amplitude as the coefficient of the series in the source. 
 
 In the unbroken $\phi^4$ theory,
\begin{equation}\nonumber
\LL=\frac{1}{2}(\partial\phi)^2-\frac{1}{2}m^2\phi^2-\frac{1}{4!}\lambda\phi^4,
\end{equation}
the classical solution is,
\begin{equation}\nonumber
\phi_{cl}=\frac{z}{1-(\lambda/48 m^2)z^2},\quad\quad\boxed{\text{unbroken sym.}}
\end{equation}
where $z$ is proportional to the source. The amplitudes are found by differentiating with respect to $z$ and setting $z=0$

\beq\label{eq:ampfromgenfun}
A_n=\braket{n|\phi_{cl}|0}=\frac{d^n\phi_{cl}}{dz^n}\Big|_{z=1}.
\eeq

We find,
\begin{equation}\label{eq:Browntreeunbroken}
A_n^{tree}=n!\Big(\frac{\lambda}{48 m^2}\Big)^{\frac{n-1}{2}},\quad\quad\boxed{\text{unbroken sym.}}
\end{equation}

with $n=3,5,\dots$. Consequently, the cross section, once multiplied by $1/n!$ for accounting for the $n$ identical bosons in the final state, will grow factorially.
 
 In the broken theory\footnote{The solution is not unique, one also can find that
  \begin{equation}\nonumber
\phi_{cl}=\phi_0\frac{1+z/2\phi_0}{1-z/2\phi_0},
\end{equation}
with $\phi_0=\sqrt{3!m^2/\lambda}$ works~\cite{Brown:1992ay}.}, one finds~\cite{Smith:1992rq},
  \begin{equation}\nonumber
\phi_{cl}=\frac{z}{1-z\sqrt{\lambda/12m^2}}.\quad\quad\boxed{\text{broken sym.}}
\end{equation}
 Here the amplitude will be
 \begin{equation}\label{eq:Browntreebroken}
A_n^{tree}=n!\Big(\frac{\lambda}{24m^2}\Big)^{\frac{n-1}{2}}.\quad\quad\boxed{\text{broken sym.}}
\end{equation}
 Again, with a factorial growth. 
 
 Loop correction does not alleviate the problem. In~\cite{Smith:1992rq, Voloshin:1994yp}, using Brown's method, and in~\cite{Argyres:1993wz} by using recursion relation method, the one loop diagrams had been calculated \footnote{For a review, see appendix~\ref{app:loopPhi36d}, where we have calculated the loop correction for the cubic theory in six spacetime dimensions.}. In both cases one needs to choose the counter-terms appropriately, otherwise no analytic solution can be found. For the unbroken theory, the leading $n$ dependence of generating function has been found to the first order in perturbation theory to be
 
 \begin{equation}\nonumber
 \phi_{cl}=\frac{z}{1-(\lambda/48 m^2)z^2}\Big(1-\frac{\lambda}{4}B\frac{(\lambda/48m^2)^2z^4}{(1-(\lambda/48m^2)z^2)^2}\Big),\quad\quad\boxed{\text{unbroken sym.}}
 \end{equation}
 
 where $B$ is given by 
 \begin{equation}\nonumber
 B=\frac{\sqrt{3}}{2\pi^2}\Big(\ln \frac{2+\sqrt{3}}{2-\sqrt{3}} -i\pi\Big).
 \end{equation}
 The amplitude is modified to
 \beq\nonumber
 A_n=A_n^{tree}\Big(1-\frac{\lambda B}{32}(3-2n+n^2)\Big).\quad\quad\boxed{\text{unbroken sym.}}
 \eeq
 
 The fact that the loop correction vanishes for $n=3$ is the result of the subtraction scheme used. 
 
 For the broken symmetry theory, the generating function is as follow
 \begin{align}\nonumber
 \phi_{cl}=\frac{z}{1-z\sqrt{\lambda/12m^2}}\Big(1-\frac{\lambda^{3/2}z}{48\pi m(1-z\sqrt{\lambda/12m^2})^2}\Big).\quad\quad\boxed{\text{broken sym.}}
 \end{align}

\subsection{Exponentiation and The \emph{Holly Grail} Function}
Later on, from analyzing the singularities of the generating function for higher loop order corrections, it was shown that they exponentiate in the $\lambda n < 1$ limit~\cite{Libanov:1994ug}. If we write $A_n=A_n^{\text{tree}}A_n^{\text{loop}}$ with $A_n^{\text{tree}}$ given in \eqref{eq:Browntreeunbroken} and \eqref{eq:Browntreebroken}, we have
 \begin{align*}
A_n^{\text{loop}}=&\exp\Big[ -\lambda n^2\frac{1}{64\pi^2}\Big( \ln(7 +4\sqrt{3})-i\pi\Big)\Big],&\quad\quad\boxed{\text{unbroken sym.}}\\
A_n^{\text{loop}}=&\exp\Big[ \lambda n^2\frac{\sqrt{3}}{8\pi}\Big].&\quad\quad\boxed{\text{broken sym.}}
\end{align*}

These exponents sum all the loops contributions to all the loop orders. This had led to the conclusion that the cross section can be written as the exponential of a function of $\lambda n$, dubbed the Holy-Grail function,
\beq\label{hollygrail}
\sigma_n\propto e^{F(\lambda n)/\lambda},
\eeq
where the Holy-Grail function, $F$, is
\begin{align}\label{holygrailunbroken}
F=&\lambda n\log \frac{\lambda n}{48}-\lambda n-(\lambda n)^2 \frac{2Re[B]}{32\pi^2}+\dots,&\quad\quad\boxed{\text{unbroken sym.}}\\\label{holygrailbroken}
F=&\lambda n\log \frac{\lambda n}{24}-\lambda n+(\lambda n)^2 \frac{\sqrt{3}}{4\pi}+\dots.&\quad\quad\boxed{\text{broken sym.}}
\end{align}

 This is a series in $\lambda n$. For $\lambda n \approx 1$, the series will blow up and we need to consider non-perturbative approaches which we shall discuss in the next sections\footnote{One can hope that summing to all order in $\lambda n$ gives a falling exponential in $n$ and this will avoid a unitarity violation. It is interesting to note that in the case of $\phi^6$ theory, it has been shown that the loop summation leads to a hyperbolic cosine~\cite{Schenk:2021yea,Khoze:2022fbf}. Hence, unitarity might not hold in this case even if we were able to sum to all order in $\lambda n$}. 

The threshold limit corresponds to the limit where the kinetic energy of the final particles vanishes. The cross sections at the threshold actually is zero, since there are no phase-space available. To be able to find the cross section slightly away from threshold, we work in the approximation where all the final particles have the same average kinetic energy,
\beq\nonumber
\epsilon=\frac{E-n m}{n m},
\eeq
where $E$ is the energy of the incoming off-shell particle. 

Since the amplitudes are spacetime independent, the cross section is approximately give by
\beq\nonumber
\sigma_n\approx \frac{1}{n!}|A_{n}|^2\Big(\frac{\epsilon^{3/2}}{3\pi}\Big)^n.
\eeq
As for the amplitudes, the energy dependence is found by recursion relations. For small energies one finds that~\cite{Libanov:1997nt}
\beq\nonumber
A_n=A_n^{tree}e^{-\frac56 n\epsilon}.
\eeq
Integrating the square of the amplitude over the phase-space near threshold, the total contribution to the Holy-Grail functions will be
\begin{align}\nonumber
F=&F_0+f(\epsilon)\lambda n,\\\nonumber
f(\epsilon)=&\frac{3}{2}(\log \epsilon/3\pi +1)-\frac{17}{12}\epsilon,
\end{align}
where $F_0$ is given by \eqref{holygrailunbroken} and \eqref{holygrailbroken}. These expressions are valid for $\lambda n <1$ and $\epsilon\ll 1$.

\subsection{Semi-Classical and Non-Perturbative Methods}
There has been considerable number of attempts to address the multi-scalar problem in 90s using different methods. For completeness, let us briefly discuss some of them and refer to the original works for further details.
\begin{itemize}
\item The \emph{Coherent State Formalism} approach is based on the steepest decent method using coherent states in QFT and is similar to Landau WKB method in quantum mechanics. The application to multi-particle had been pioneered by Son~\cite{Son:1995wz} and more recently studied in more detail in~\cite{Khoze:2018mey} and \cite{Demidov:2022ljh}. In \cite{Khoze:2018mey}  the loop corrections in the $\lambda n\gg1$ was found to be
\beq\label{holygrailnonperturb}
F_0=\lambda n \log \lambda n -\lambda n + 0.85(\lambda n)^{3/2},
\eeq
in the broken symmetry $\phi^4_{\text{4D}}$.  This formula is the basis for the Higgsplosion hypothesis (see section~\ref{sec:multihiggs}).

We can add to this the energy dependence in the $\epsilon \ll1$ limit, in the next to leading order~\cite{Libanov:1997nt}, which is 
\beq\nonumber
f(\epsilon)=\frac32\Big(\log\epsilon/3\pi+1\Big)-\frac{17}{12}\epsilon+\frac{1327-96\pi^2}{432}\epsilon^2.
\eeq

For the unbroken $\phi^4_{\text{4D}}$, in \cite{Demidov:2022ljh}, the method has been extended to the $\lambda n \gg 1$ and $\epsilon \le \text{few} \times m$. By fitting to the numerical calculation, they found,
\beq\label{eq:Demidov2022}
F=-\frac{3}{4}\log [(d_1 m/\epsilon)^2+d_2]\lambda n+g_{\infty}(\epsilon)\lambda n,
\eeq
where $d_i\approx [10.7, 30.7]$, and $g_{\infty}(\epsilon)$ is negative. 

\item The \emph{Lipatov Method} is based on analytically continuing to the negative values of $\lambda=-\lambda'$, where the potential will be inverted and all the amplitudes will acquire an imaginary part. One can write the real part of the amplitude for positive $\lambda$ in terms of the imaginary part for the negative $\lambda$, through the dispersion relation
\beq\label{eq:amplitudelipatov}
A_n(p_i,\lambda)=const. +\frac{\lambda}{\pi}\int_0^{\infty}d\lambda'\frac{Im[A_n(p_i,\lambda')]}{(\lambda'+\lambda)\lambda'}.
\eeq
For negative $\lambda$, the imaginary part of the amplitudes can be calculated using the instanton solution of the inverted potential~\cite{Lipatov:1977hj}. The authors of~\cite{maggshif92, maggshif91} have applied this idea to the amplitude $A_n(2\to n-2)$, in the scalar theory. Writing the amplitude as
\beq\nonumber
A_{n}=\sum_l a^{l}_n\lambda^{(n-1)/2+l},
\eeq 
where $l$ is the loop order ($l=0$ is the tree level). The coefficients $a_n^l$ can be calculated by expanding \eqref{eq:amplitudelipatov}:
\beq\nonumber
a_n^l=(-1)^{n/2+l}\frac{1}{\pi}\int d\lambda'\frac{Im[A_n(p_i,\lambda')]}{\lambda'^{n/2+l}}.
\eeq 

 It is shown that the factorial behavior shows up in this method as well:
\begin{align}\nonumber
a^l_n\propto&(l+n/2)!, \quad\quad & l\gg1, n=\OO(1),\\\nonumber
a^l_n\propto&\Gamma(l+n+3/2), \quad\quad & n \gg1.
\end{align}


\item Yet another way is to use \emph{Functional Schrodinger Method}. The most important outcome for our purpose is that the cross section should not grow at high multiplicity of the final particles. For example, it can be shown that the amplitude for $n\lambda\gg 1$, the cross section decays exponentially\footnote{This can also be shown in a quantum mechanical system~\cite{Jaeckel:2018ipq,Jaeckel:2018tdj}.}~\cite{cornwall92, Cornwall:1993rh}
\begin{equation}\nonumber
\sigma_n\propto\text{exp}(-\frac{\pi}{2}n).
\end{equation}
\end{itemize}

\subsection{Enhancement of Multi-Higgs cross section via Higgsplosion}\label{sec:multihiggs}

Finally, let us review the most recent work on the \emph{multi-Higgs} cross section~\cite{Khoze:2014kka, Khoze:2015yba, Degrande:2016oan, Khoze:2017tjt, Khoze:2017ifq}. For observing the factorial growth in the experiment, one needs to find the energy dependence away from threshold limit. None of the method known gives a good approximation in large $\epsilon$ limit, that is when the final particles are relativistic. It is, however, possible to extract the epsilon dependence at tree level using Monte Carlo simulation, if we accept the ansatz that the cross section will exponentiate into a Holy-Grail form \eqref{hollygrail}, specifically that the dependence on $\lambda$ comes in $\lambda n$ form\cite{Khoze:2015yba}. 

\begin{figure}[ht!]
\begin{center}
\includegraphics[width=0.8\textwidth]{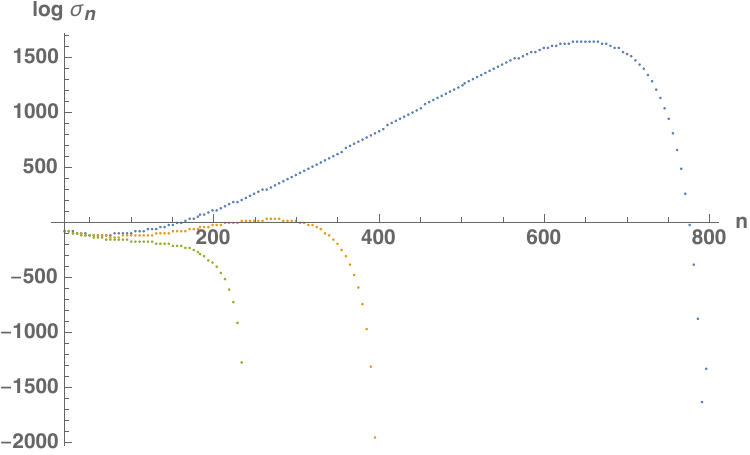}
\end{center}
\caption{The multi-Higgs phenomena. From top to bottom, the center-of-mass energy is $100, 50, 10$ TeV. The suppression at the far right is due to the phase-space, where particles are non-relativistic.}
\label{fig:khoze}
\end{figure}

The point is that, at tree level, the $f(\epsilon)$ does not depend on $\lambda$, and since the expansion parameter is $\lambda n$, it will not depend on $n$ either. Hence, if we look at the ratio of two consecutive cross sections, we find that
\beq\nonumber
\log \sigma_{n+1}/\sigma_n=F_0(n+1)/\lambda-F_0(n)/\lambda+f(\epsilon).
\eeq 
Since $F_0$ is known, the authors of~\cite{Khoze:2015yba} used Madgraph~\cite{Alwall:2014hca} to find a fit for $f(\epsilon)$ for $n=5$. We have not redo their simulation; instead used a fit to their graph in figure 2 of~\cite{Khoze:2015yba} to display their result here. We used the fit together with the expression in \eqref{holygrailnonperturb} for the broken theory (note that we do not have a non-perturbative equation for the unbroken theory), to produce Fig.~\ref{fig:khoze} (compare to figure 6 in~\cite{Khoze:2014kka}; however, in this paper the authors used the perturbation result for the loop correction \eqref{holygrailbroken} which is not valid at large $\lambda n$). We can see that the cross section becomes un-suppressed at finite values of $\epsilon$ before they fall down due to suppression from the phase-space. 

It is important to note that the enhancement of the cross section at $\mathcal{O}(100)$ TeV energies and low number of final particles (what is suggested to look for in experiments) is not so much a result of the factorial growth but the fact that the correction to the tree result had come with a positive sign in \eqref{holygrailnonperturb}. If we naively just use the tree result, we would have an enhancement at much higher energies $\approx\mathcal{O}(10^4)$ TeV.

Furthermore, the cross section becomes un-suppressed at values of $\epsilon$ for which the final particles are relativistic. As explained earlier, we will use jet generating function method to sum the contribution for producing $n$ jets to study this region of phase-space.

\section{Jet Generating Functional Method}\label{sec:jetGF}

Jet generating functional (JGF)~\cite{Ellis:1991qj, Dokshitzer:1991wu} is a functional of a probing function, $u(p_i)$, where $p_i$ is the momentum of $i$'th particle, such that when expanding in $u$, the coefficients are the exclusive particle cross sections. Here, we will use the simplified version where the coefficients are jet rate and we have a function instead of a functional. This function can be find from the original functional, if it was available, by integrating over the momentums of the particles to obtain jets.

Calling the generating function $\Phi$, by definition the jet rates are 
\beq\label{eq:jetrate}
R_n\equiv\frac{\sigma_{n-\text{jets}}}{\sigma_{\text{total}}}\equiv\frac{1}{n!}\Big(\frac{\partial}{\partial u}\Big)^n\Phi[u,t]\Big|_{u=0},
\eeq
where $t$ is the energy of the incoming particle that produces jets.

One can also use this function to find the average multiplicity number as follow
\beq\label{eq:jetaveragen}
\bar{n}=\sum_n nR_n=\frac{\partial}{\partial u}\Phi[u,t]\Big|_{u=1}.
\eeq

 Chang and Lau~\cite{chang:1977}, and later Taylor~\cite{Taylor:1978hu}, were the first ones\footnote{As far as we checked.} to apply the idea to $\phi^3$ theory in six spacetime dimensions. Later, Kinoshi and others~\cite{Konishi:1979cb} used Altarelli-Parisi evolution equation~\cite{Altarelli:1977zs,Gribov:1972, Lipatov:1974qm, Dokshitzer:1977} to ``derive" an equation for gluon and quark JGF, mainly based on jet evolution from a perturbation understanding. In our understanding, there is no real derivation of the generating functional from first principles. One usually starts by writing a probabilistic equation for it.
 
Let us work with a cubic theory, whether QED or QCD, or $\phi^3$ theory. The equation for a JGF can be written as follow

\begin{align}
\Phi[u,t]=&\Phi[u,t_0]\Delta(t,t_0)\nonumber\\
&+\int _{t_0}^t dt'dz\PP(z,t')\frac{\Delta(t,t_0)}{\Delta(t',t_0)}\Phi[u,z^2t']\Phi[u,(1-z)^2t']\label{eq:jetGFgeneral}.
\end{align}
where the $z$ is the energy fraction of one of the daughter particles. This equation gives how a particle at scale $t$ evolves and splits into other particles. A jet with a large energy wants to split into more jets unless its energy is of order $t_0$, where by definition it cannot split anymore: $\Phi(u,t_0)=u$. The first term on the right hand side describes a particle if it had not split into other jets. The $\Delta(t,t_0)$, hence, is the probability of not splitting given the two scales, $t_0$, and $t$. It is called the Sudakov form factor~\cite{Sudakov:1954sw, collins:1989} . The second term on the right hand side is the sum (turned into an integral) of probabilities of the particle when it splits into two other jets at scale $t'$. The $\PP(z,t)$, called Altarelli-Parisi function is the probability density of splitting into two particles with energy fractions $z$ and $1-z$. The fraction $\Delta(t,t_0)/\Delta(t',t_0)\approx\Delta(t,t')$ is the probability that the particle does not split before the scale $t'$.

Knowing the Altarelli-Parisi, we can form the Sudakov form factor~\cite{Schwartz:2013pla}. First note that the probability of splitting between two scales, $t_0$ and $t$ is
\beq\nonumber
R(t,t_0)=\int_0^1dz\int_{t_0}^tdt'\PP(z,t').
\eeq
 Sudakov factor, by definition, is the probability of not splitting. Evaluating it for scales $t_0$ and $t+\delta t$, where $\delta t$ is infinitely small, gives
\begin{align}\nonumber
\Delta(t+\delta t,t_0)=&\Delta(t,t_0)\bigg(1-\int_0^1dz\int_{t}^{t+\delta t}dt'\PP(z,t')\bigg),\\\nonumber
\approx&\Delta(t,t_0)-\Delta(t,t_0)\delta t\int dz \PP(z,t).
\end{align}
On the other hand,
\beq\nonumber
\Delta(t+\delta t,t_0)=&\Delta(t,t_0)+\delta t\frac{d}{dt}\Delta(t,t_0). 
\eeq
Hence, we find the differential equation
\beq\nonumber
\frac{d}{dt}\Delta(t,t_0)=\Delta(t,t_0)\int dz \PP(z,t),
\eeq
with $\delta(t_0,t_0)=1$ by definition, the solution is
\beq\label{eq:Sudakovfactorequation}
\Delta(t,t_0)=\exp\Big[-\int_0^1dz\int_{t_0}^tdt'\PP(z,t')\Big].
\eeq

The variable $t$ is usually taken to be,
\beq\nonumber
t=\frac{k_T^2}{z^2(1-z)^2},
\eeq

where $K_T$ is the transverse momentum of daughter jets. For a two-body decay, it is
\beq\label{eq:variablet}
t=Q^2(1-\cos^2 \theta),
\eeq
where the angle is between the decayed particles. Whether $t$ is a good variable or not is a matter of comparing to experiment. It is claimed that the generating function that satisfies \eqref{eq:jetGFgeneral}, with the so defined $t$, correctly sums the divergent logs for jets in $K_T$-algorithms~\cite{Catani:1991hj}.

In QCD, equation \eqref{eq:jetGFgeneral} gives the jet rate which phenomenologically gives different scaling for abelian and non-abelian splitting: Poisson pattern and staircase pattern respectively, that can be examined in experiment~\cite{Plehn:2015dqa}. 

Our main goals would be to find $\PP$ and $\Delta$, and plug them into \eqref{eq:jetGFgeneral} and use \eqref{eq:jetrate} to find the jet rates.

\section{\texorpdfstring{$\phi^3$}{phi3} Theory In Six Spacetime Dimensions}\label{sec:phi36D}
The cubic scalar theory in six dimensions provides a good example for analyzing the multi-scalar problem. Different bits of the jet calculations we want to do had been done before, which are a good check. Unlike in four dimensions, the coupling is marginal. And unlike quartic theory, particles can split into two particles. These two facts makes finding the splitting function and the Sudakov factor easier.

Let us first briefly discuss the threshold limit of the multi-scalar production. 

The Lagrangian is
\begin{equation}\nonumber
\mathcal{L}=\frac{1}{2}(\partial\phi)^2-m^2\phi^2-\frac{g}{3!}\phi^3+\frac12\delta_Z(\partial\phi)^2-\frac12 \delta_m\phi^2-\frac{1}{3!}\delta_{g}\phi^3-\delta_{\tau}\phi.
\end{equation}

The multi-particle amplitude via the classical generating function method of Brown is basically counting the number of tree graphs and does not depend on the spacetime dimension. In~\cite{Papadopoulos:1993aw} it has been calculated for a general $\phi^m$ interaction. For cubic theory one finds

\begin{equation}\nonumber
\phi_{0}=\frac{z}{(1-\frac{g}{12}z)^{2}},
\end{equation} 

for $m^2=1/2$. This gives the amplitude at tree level \eqref{eq:ampfromgenfun},
\begin{equation}\nonumber
A_{n}=n n!\Bigg(\frac{g}{12}\Bigg)^{n-1}.
\end{equation}

Again, we see a factorial growth.

 We have calculated the loop correction in appendix~\ref{app:loopPhi36d}. The result is

\beq\nonumber
\phi_1(z)=\frac{18g^2z}{(1-\frac{g}{12}z)^4},
\eeq
which gives the amplitude at one loop level,
\begin{align}\label{eq:ampbrownphi36d}
A_n=\frac{d^n}{dz^n}(\phi_0+\phi_1)\big|_{z=0}=nn!\Big(\frac{g}{12}\Big)^{n-1}\Big[1+3g^2(n^2+3n+2) \Big].
\end{align}
Interestingly, we find that the conjecture that the expansion is in $g n$, holds here as well (although, here the first term is $(gn)^2$).

The cross section,
\beq\label{eq:xsecphi36d}
\sigma_n\propto \frac{|A_n|^2}{n!}\times (\text{Phase Space}),
\eeq

again suffers from a factorial growth. Since we are interested in whether the factorial is present for jet cross sections or not, we will skip finding the energy dependence near threshold and analyzing this model with other non-perturbative methods explained in the previous section.

 As stated previously, our main goal is to find the AP function and the Sudakov form factor. In order to do so, it is instructive to check the cancelation of the mass-divergence. Under Bloch-Nordsieck theorem~\cite{Bloch:1937pw} and Kinoshita-Lee-Nauenberg theorem~\cite{Kinoshita:1962ur}, in QED and QCD respectively, the mass singularities cancel in the total cross section between the virtual contribution and the radiation contribution~\cite{Peskin:1995ev, Schwartz:2013pla}. This phenomenon happens for the scalar theories too. Here, we will follow Srednicki~\cite{Srednicki:2007qs} to show the cancelation in $2\phi\to2\phi$ scattering, Fig.~\ref{2phoTO2phiinPhi36D}.
 \begin{figure}[ht!]
   \begin{center}
  \centering
\begin{fmffile}{pp_ppPHI3}
    \begin{fmfgraph}(150, 65)
     \fmfleft{l1,l2}
     \fmfright{r1,r2}
     \fmf{dashes}{l2,o}
     \fmf{dashes}{l1,o}
     \fmf{dashes}{o,r1}
     \fmf{dashes}{o,r2}
          \fmffreeze
    \fmfblob{30}{o}
    \end{fmfgraph}
    \end{fmffile}
  
    \caption{$\phi\phi\to \phi\phi$.}
    \label{2phoTO2phiinPhi36D}
      \end{center}
  \end{figure}

 \begin{figure}[ht!]
   \centering
{\begin{fmffile}{hhh_Loop00}
    \begin{fmfgraph}(130, 65)
     \fmfleft{l1,l2,l3}
     \fmfright{r1,r2,r3}
     \fmf{dashes}{l2,ol1}
     \fmf{dashes}{ol1,or1}
     \fmf{dashes}{ol1,or3}
     \fmf{dashes}{or3,r3}
     \fmf{dashes}{or1,r1}
     \fmffreeze
     \fmf{dashes}{or1,or3}
    \end{fmfgraph}\end{fmffile}
  }
    \caption{Vertex correction.}
    \label{fig:phi3vertexcorrection}
  \end{figure}
Following Srednicki, in the $m\to 0$ limit, we use the $\bar{\text{MS}}$ scheme. In this scheme, and in the mass-less limit, the 1PI loop vertex corrections, Fig.~\ref{fig:phi3vertexcorrection}, do not depend on the mass after renormalization\footnote{\label{footnote:scheme}In other schemes, the counter term is added such that $F_3(p^2=0)=g$. In this case, the counter term will have an IR divergence as well once we take $m\to 0$ limit.}. The leading and next to leading order read as, 
\beq\nonumber
\mathcal{M}=R^2\mathcal{M}_0\Big[1-\frac{11}{12}\frac{g^2}{(4\pi)^3}\Big(\log\frac{s}{\mu^2}+\OO(m)\Big)+\dots\Big],
\eeq
where $\mathcal{M}_0=g(s^{-1}+t^{-1}+u^{-1})$ is the tree level amplitude. However, and it will be the case in the other two theories explained in the next sections, the residue, $R$, of the two-point function has an IR divergence. At the mass pole it is,
\beq\label{eq:Rphi36d}
R=\frac{1}{1-\Pi'(m_{phy})}\approx\frac{1}{1-\Pi'(m)}\approx 1-\frac{1}{12}\frac{g^2}{(4\pi)^3}\log\frac{\mu^2}{m^2},
\eeq 
where prime is differentiation with respect to $p^2$. We find,
\beq\nonumber
\mathcal{M}=\mathcal{M}_0\Big[1-\frac{g^2}{(4\pi)^3}\Big(\frac{11}{12}\log\frac{s}{\mu^2}+\frac{1}{6}\log\frac{\mu^2}{m^2}+\OO(m)\Big)+\dots\Big].
\eeq
The divergent log in $m\to 0$ limit would not sum by renormalization of the coupling constant. In order to be able to take the mass-less limit, we add to $\sigma_{2\to 2}$ cross section the radiations off of the legs. The radiation from a single is shown in Fig. \ref{fdiag:phi3s6dplit}. If we separate its contribution to the cross section, we find.
\beq\label{eq:split2}
\frac{1}{2!}\int \frac{g^2}{(p_a^2-m^2)^2}d^6p_bd^6p_c,
\eeq
where line $a$ has splatted to $b$ and $c$. A delta function,
\beq\nonumber
1=\int d^6p_a 2E_a (2\pi)^5\delta^5(\vec{p}_a-\vec{p}_b-\vec{p}_c),
\eeq
 is added to turn one of the phase-space integrals into the integral over $p_a$, so that we can factor out the cross section of one less particle process. 
 \begin{figure}[ht!]
   \centering
\begin{fmffile}{h_to_hh0}
    \begin{fmfgraph*}(150, 85)
     \fmfleft{l}
     \fmfright{r1,r2}
     \fmf{dashes,tension=2.5,label=$a$}{l,o}
     \fmf{dashes,label=$b$}{o,r2}
     \fmf{dashes,label=$c$}{o,r1}
    \end{fmfgraph*}\end{fmffile}
\caption{$\phi\to \phi\phi$. Momentums are flowing to the right.}
  \label{fdiag:phi3s6dplit}
  \end{figure}
  
Writing the integral in terms of the angel between the two, $\theta$, and the energy fraction of one of the daughter particles, $z=E_b/E_c$, and using the assumptions $m^2\ll Q^2$ and $m^2/Q^2< \delta^2$, where $\delta$ is the resolution angle, we find
\beq\label{eq:radiationphi36d}
\frac{g^2}{(4\pi)^3}\int_0^1 z(1-z)dz\int_0^{\delta}\frac{\theta^3d\theta}{(\theta^2+\frac{m^2}{E_a^2}f(z))^2}=\frac{g^2}{12(4\pi)^3}\log \frac{\delta^2 E_a^2}{m^2}+\dots,
\eeq
where $f(z)=(1-z+z^2)/(z-z^2)^2$.
There is no soft divergence, just the collinear divergence. Hence, the single log. This is just what we need for canceling the log from the virtual diagrams and turning it into a log of $\delta$. Using $E_a^2=s/4$, we find
\begin{align}\nonumber
\sigma_{2\to 2}=&\Big(1+\frac{g^2}{12(4\pi)^3}\log \frac{\delta^2 E_a^2}{m^2}\Big)^4|\mathcal{M}|^2,\\\nonumber
=&|\mathcal{M}_0|^2\Big[1-\frac{g^2}{(4\pi)^3}\Big(\frac{3}{2}\log\frac{s}{\mu^2}+\frac{1}{3}\log\frac{1}{\delta^2}+\OO(m)\Big)+\dots\Big].
\end{align}

The $\frac{g^2}{3(4\pi)^3}\log \delta^2$ can be interpreted as the probability of the four legs not splitting. The fact that it blows up, is a signal that we need to re-sum the perturbation series. This is usual practice in QED and QCD, and can be done for cubic scalar in six dimensions too, by summing further radiations. The resumed probability of a leg not-splitting, called the Sudakov form factor, is
\beq\nonumber
\Delta=\exp\Big[-\frac{g^2}{12(4\pi)^3}\log[1/\delta^2]\Big].
\eeq
 The cross section becomes
\beq\label{eq:twototwocrosssectionphi36d}
\sigma_{2\to 2}= \Delta^4|\mathcal{M}_0|^2\Big[1-\frac{g^2}{(4\pi)^3}\Big(\frac{3}{2}\log\frac{s}{\mu^2}+\OO(m)\Big)+\dots\Big].
\eeq
The remaining logs are removed by renormalization. 

Let us note that these results are correct for $\frac{m^2}{Q^2}<\delta^2$. What if $\frac{m^2}{Q^2}>\delta^2$? In this limit, the integral in \eqref{eq:radiationphi36d} vanishes as $\delta\to 0$. 

What happens is that in the massless limit, when $\frac{m^2}{Q^2}<\delta^2$, the Sudakov form factor sums the radiation contribution and virtual corrections in such a way to give the probability of not-splitting, given some jet definition (loosely, we can change $\delta$ with other jet definition, for example jet mass or trust). Now, for $\delta<m/Q$, the Sudakov factor sums only the virtual logs, 
\beq\nonumber
\Delta=\exp\Big[-\frac{g^2}{12(4\pi)^3}\log\frac{Q^2}{m^2}\Big].
\eeq
This can be shown in Soft-Collinear Effective Theory~\cite{Becher:2014oda}, that in a massive theory the Sudakov form factor sums the IR divergent logs of virtual diagrams. In other words, we can think of the mass as the IR regulator.

Let us go back to the integral in \eqref{eq:radiationphi36d}. Integrating over the range of $\theta$ where the opening angle is bigger than the resolution, gives the probability of a line splitting. Hence, we can read the AP function. From equation \eqref{eq:variablet}, we find,
\beq\label{eq:APfunctionPhi36d}
\mathcal{P}(z,t)=\frac{g^2}{(4\pi)^3}\frac{z(1-z)}{2t}.
\eeq


 So far, we have neglected the running of the coupling, since it did not play a role in the cancelation of the IR divergences. Let us evaluate the Sudakov form factor, taking into account the running. In the cubic theory in six dimensions, to the first order, we have
\beq\label{eq:runninginphi36d}
g^2(t)=\frac{g(\mu^2)^2}{1+\frac{3}{4}\frac{g(\mu^2)^2}{(4\pi)^3}\log t/\mu^2}.
\eeq
 To find the Sudakov factor \eqref{eq:Sudakovfactorequation}, we need to evaluate
 \begin{align}\nonumber
 \int_0^1 dz\int_{t_0}^{t} dt' \mathcal{P}(z,t')&= \int_0^1 \frac{z(1-z)dz}{2}\int_{t_0}^{t} dt' \frac{g^2(t')}{(4\pi)^3}\frac{dt'}{t'}\\\nonumber
&=-\frac{1}{12}\int_{t_0}^{t}\frac{-4 d g^2}{3g^2}\\\nonumber
&=\frac{1}{9}\log \frac{g^2(t)}{g^2(t_0)}.
 \end{align}
where we have changed the variable to $g^2$, using the equation \eqref{eq:runninginphi36d}. Hence, the Sudakov factor \eqref{eq:Sudakovfactorequation}, is
\beq\label{eq:Sudakovfactorphi36d}
\Delta(t,t_0)=\bigg(\frac{g^2(t)}{g^2(t_0)}\bigg)^{\frac{1}{9}}.
\eeq
 We now have all the ingredients for writing the differential equation for the generating function. For the cubic theory it is \eqref{eq:jetGFgeneral}, 
\begin{align}
\Phi_{(3)}[t]=&u\Delta(t,t_0)\nonumber\\
&+\int_{t_0}^t dt'\frac{\Delta(t,t_0)}{\Delta(t',t_0)}\int dz\mathcal{P}(z,t')\Phi_{(3)}[z^2t']\Phi_{(3)}[(1-z)^2t']\label{jgf6d}.
\end{align}

Noting that the $z$ dependence of the functions under integral are logarithmic (through the Sudakov factors and the coupling), we discard the $z$-dependence of these functions for simplicity. The simplified equation becomes,
\begin{align}\nonumber
\Phi_{(3)}[t]=&u\Delta(t,t_0)\\\nonumber
&+\int_{t_0}^t dt'\frac{\Delta(t,t0)}{\Delta(t',t_0)}\int dz\mathcal{P}(z,t')\Phi_{(3)}[t']\Phi_{(3)}[t'].
\end{align}

Now, differentiation with respect to $t$ from both sides, and using the formula for the Sudakov factor \eqref{eq:Sudakovfactorequation}, gives
\beq\nonumber
\frac{d\Phi_{(3)}[t]}{dt}=\frac{1}{\Delta}\frac{d\Delta}{dt}\Big(\Phi_{(3)}[t]-\Phi_{(3)}^2[t]\Big).
\eeq

Rearranging the equation and integrating both sides, and using the boundary condition, $\Phi[t_0]=u$, we find\footnote{This also is the JGF for a gluon in pure Yang-Mills theory, that is without splitting of the quark and anti-quarks into gluons and vice versa, which leads to \emph{staircase} pattern for the gluons~\cite{Plehn:2015dqa} and had been checked in experiment.},
\beq\label{phi36dGF}
\Phi_{(3)}[t]=\frac{u}{u+(1-u)\Delta^{-1}}.
\eeq

This is in agreement with Taylor~\cite{Taylor:1978hu} and Kinoshi~\cite{Konishi:1979cb}.

Using \eqref{phi36dGF} and \eqref{eq:jetrate}, we find that
\beq\label{eq:Rnph36d}
R_n=\Delta(1-\Delta)^{n-1}.
\eeq

Using \eqref{eq:jetaveragen}, we can find the average jet multiplicity
\beq\nonumber
\bar{n}=\frac{\partial}{\partial u}\Phi\Big|_{u=1}=\Delta^{-1}.
\eeq

Plugging in the Sudakov factor \eqref{eq:Sudakovfactorphi36d}, we find,
\begin{align}
R_n&=\bigg(\frac{g^2(t)}{g^2(t_0)}\bigg)^{\frac{1}{9}}\bigg(1-\Bigg(\frac{g^2(t)}{g^2(t_0)}\bigg)^{\frac{1}{9}}\Bigg)^{n-1}\label{eq:jetratephi36d},\\\nonumber
\bar{n}&=\bigg(\frac{g^2(t_0)}{g^2(t)}\bigg)^{\frac{1}{9}}.
\end{align}
As $g(\mu^2)\to 0$, we have $\bar{n}\to 1$, which means that there are no splitting. The non-integer power means that we cannot find this expression from perturbation theory. In order to compare to fix order result, we expand \eqref{eq:jetratephi36d} for small $g_{\mu}$. We find,
\beq\label{eq:ratephi36dexpand-1}
R_n=\bigg(\frac{g(\mu^2)^2}{12(4\pi)^3}\bigg)^{n-1}\bigg(\log \frac{t}{t_0}\bigg)^{n-1}\bigg(1-\frac{g(\mu^2)^2}{12(4\pi)^3}\log\frac{t}{t_0}\bigg).
\eeq

As we will explain in section~\ref{sec:offsh}, this is not the correct expression to compare to the equation~\eqref{eq:xsecphi36d}. However, we can see that there is no factorial growth in the jet-cross sections. We should keep in mind that the amplitude we found in \eqref{eq:ampbrownphi36d} was near threshold, while the jet-cross sections is a sum of many cross sections with collinear particles, and all very energetic.

\section{\texorpdfstring{$\phi^4$}{phi4} Theory In Four Spacetime Dimensions} \label{sec:phi44D}

We now turn to the $\phi^4$ theory
\begin{equation}\nonumber
\LL=\frac{1}{2}\big(\partial\phi\big)^2-\frac12m^2\phi^2-\frac{\lambda}{4!}\phi^4.
\end{equation}
 
 As in the previous section, we start by checking the cancelation of the mass-divergence in $\phi\phi\to\phi\phi$ scattering, Fig.~\ref{2phoTO2phiinPhi36D}. We first note that the IR divergence should appear at two loops order. This is because the contribution of a single radiation $\phi\to3\phi$ to the total cross section is at $\lambda^4$ order. If there were an IR divergence at one loop order, there would be a $\lambda^3$ contribution, from the interference of the virtual diagram with the tree diagram, to the total cross section.

 \begin{figure}[ht!]
   \begin{center}
   \begin{eqnarray*} 
   (1)\parbox{40mm}{ 
\begin{fmffile}{pp_ppPHI4_v1}
    \begin{fmfgraph}(75, 35)
     \fmfleft{l1,l2}
     \fmfright{r1,r2}
     \fmf{dashes}{l2,ol}
     \fmf{dashes}{l1,ol}
     \fmf{phantom}{ol,o1} 
     \fmf{dashes}{o1,r1}
     \fmf{dashes}{o1,r2}
          \fmffreeze
                   \fmf{dashes,right,tension=0}{ol,o1}
     \fmf{dashes,left,tension=0}{ol,o1}
    \end{fmfgraph}
    \end{fmffile}
  }
(2) \parbox{40mm}{  
 \begin{fmffile}{pp_ppPHI4_v2}
    \begin{fmfgraph}(75, 35)
     \fmfleft{l1,l2}
     \fmfright{r1,r2}
     \fmf{dashes}{l2,ol}
     \fmf{dashes}{l1,ol}
     \fmf{phantom}{ol,o1} 
          \fmf{phantom}{o1,o2} 
     \fmf{dashes}{o2,r1}
     \fmf{dashes}{o2,r2}
          \fmffreeze
     \fmf{dashes,right,tension=0}{ol,o1}
     \fmf{dashes,left,tension=0}{ol,o1}
     \fmf{dashes,right,tension=0}{o1,o2}
     \fmf{dashes,left,tension=0}{o1,o2}
    \end{fmfgraph}
    \end{fmffile}
  }
 (3) \quad \parbox{40mm}{
\begin{fmffile}{pp_ppPHI4_v3}
    \begin{fmfgraph}(75, 35)
     \fmfleft{l1,l2,l3,l4}
     \fmfright{r1,r2,r3,r4}
     \fmf{dashes,tension=2}{l2,ol1}
      \fmf{dashes,tension=2}{l3,ol1}
     \fmf{dashes}{ol1,or1}
     \fmf{dashes}{ol1,or2}
     \fmf{dashes}{or2,r4}
     \fmf{dashes}{or1,r1}
     \fmffreeze
     \fmf{dashes,tension=.1,right=0.3}{or2,or1}
     \fmf{dashes,left=0.3,tension=0.1}{or2,or1}
    \end{fmfgraph}
    \end{fmffile}
  }
      \end{eqnarray*}    
      \end{center}
    \caption{Types of virtual corrections to $\phi\phi\to \phi\phi$.}
    \label{ph4viratualdiag}
  \end{figure}
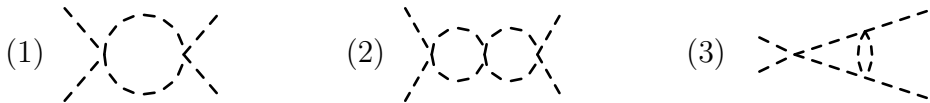

\begin{figure}[ht!]
   \begin{center}
\begin{fmffile}{sunset_PHI4t}
    \begin{fmfgraph}(150, 65)
     \fmfleft{l1}
     \fmfright{r1}
     \fmf{dashes,tension=2}{l1,o1}
      \fmf{dashes,tension=2}{r1,o2}
     \fmf{dashes}{o1,o2}
     \fmffreeze
     \fmf{dashes,tension=0,right}{o1,o2}
     \fmf{dashes,left,tension=0}{o1,o2}
    \end{fmfgraph}
    \end{fmffile}
  
    \caption{The ``sunset" diagram.}
    \label{sunset_diag}
      \end{center}
  \end{figure}
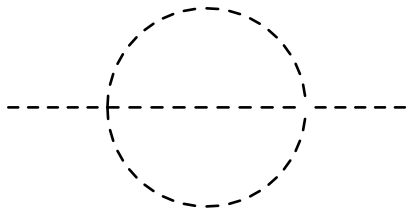

 In fact, in the $\bar{\text{MS}}$ scheme, there are no logs of mass in the virtual diagrams of Fig. \ref{ph4viratualdiag}\footnote{See footnote \ref{footnote:scheme}.}. The logs come from the contribution of the ``sunset" diagram, Fig. \ref{sunset_diag}, to the residue at the physical mass pole,
 \beq\nonumber
R=\frac{1}{1-\Pi'(p^2=m^2)}.
\eeq

 Which is 
\beq\label{eq:Rphi44d}
R\approx1+\Pi'(m^2)=1+\frac{\lambda^2}{12(4\pi)^4}\log\frac{m^2}{\mu^2}+\OO(\lambda^2)\times \text{const.}
\eeq

 Hence, the fact that IR divergences should be at $\lambda^4$, is related to the fact that there is no field renormalization in the quartic theory at one loop level. 
 
 To check this, let us write down the virtual diagrams of Fig. \ref{ph4viratualdiag}, in $\bar{\text{MS}}$ scheme, in the $m\to 0$ limit. The ``fish" diagram with the counter term added, is
\beq\nonumber
M_{(1)}=\frac{-i\lambda^2}{2(4\pi)^2}\sum_{A=s,t,u}\Big(\log\frac{A}{\mu^2}-2\Big),
\eeq
where $\mu$ is the renormalization scale and $s$, $t$, and $u$ are the Peskin parameters. The double ``fish" diagrams, is
\beq\nonumber
M_{(2)}=\frac{-i\lambda^2}{4(4\pi)^4}\sum_{A=,s,t,u}\Big(\log\frac{A}{\mu^2}-2\Big)^2.
\eeq

The ``vertex correction" diagrams is

\beq\nonumber
M_{(3)}=\frac{-i\lambda^3}{(4\pi)^4}\sum_{A=s,t,u}\Big(\frac12\log^2\frac{A}{\mu^2}-3\log\frac{A}{\mu^2}+\frac{11}{2}\Big).
\eeq

  The Logs in these expressions are absorbed into the renormalization of the coupling constant. 
  
  For the $2\phi\to2\phi$ scattering, we have a factor of $\sqrt{R}$ in the S matrix for each external leg coming from LSZ formula. We thus have, in $m\to 0$ limit, 
\beq\label{eq:rasR2M2}
\sigma_{2\to 2}= R^4|M|^2=\Big(1+\frac{\lambda^2}{3(4\pi)^4}\log\frac{m^2}{\mu^2}\Big)|M|^2.
\eeq

\begin{figure}[ht!]
   \centering
{\begin{fmffile}{h_to_hhh}
    \begin{fmfgraph*}(150, 85)
     \fmfleft{l1,l2,l3}
     \fmfright{r1,r2,r3}
     \fmf{dashes,tension=2.5,label=$a$}{l2,o1}
     \fmf{dashes,label=$c$}{o1,r2}
     \fmf{dashes,label=$d$}{o1,r1}
     \fmf{dashes,label=$b$,side=right}{o1,r3}
    \end{fmfgraph*}\end{fmffile}
  }
  \caption{$\phi\to 3\phi$ splitting. Momentums are flowing to the right. }
  \label{fig:phi4splitting}
  \end{figure}
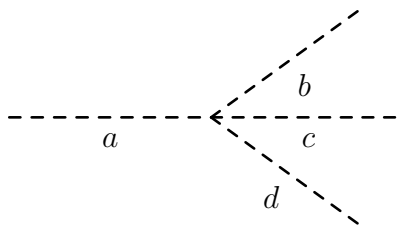

To find the physical amplitude in the $m\to 0$ limit, we need to add to \eqref{eq:rasR2M2} the radiation off of the legs. A leg, labeled $a$, splits into 3 other legs, $b$, $c$, and $d$, as shown in~Fig.~\ref{fig:phi4splitting}.

  The splitting is proportional to
  \beq\nonumber
  \frac{1}{3!}\int \frac{\lambda^2}{(p_a^2-m^2)^2}d^4p_bd^4p_cd^4p_d.
  \eeq
  We multiply this by 
\beq\nonumber
1=\int d^4p_a2 E_a(2\pi)^3\delta^3(\vec{p_a}-\vec{p_b}-\vec{p_c}-\vec{p_d}),
\eeq

and isolate the $d^4p_a$ integral to be absorbed into the cross section with two less legs. What remains is
\beq\nonumber
\frac{\lambda^2}{24(2\pi)^6}\int\frac{E_a}{E_bE_cE_d(p_a^2-m^2)^2}\sin \theta_bd\phi_b\sin\theta_cd\phi_cE_b^2E_c^2dE_bdE_c,
\eeq
where we used $E_i=|\vec{p_i}|$. We can simplify this integral by changing the variables to: $\theta, \alpha, x$, and $y$. Where $\theta$ is the angle between $b$ and $c$; $\alpha$ is the angle between $d$ and the vector sum of $b$ and $c$; and x and y are defined as follow
\begin{align*}	
E_b&=(1-x)(1-y)E_a,\\
E_c&=x(1-y)E_a,\\
E_d&=yE_a.
\end{align*}
The integral becomes
\begin{align}\label{eq:phi4splitting}
\frac{\lambda^2}{24(2\pi)^4}\int_0^1dx\int_0^1dy\int_0^{\delta}\theta d\theta\int_0^{\delta}\alpha d\alpha\frac{-x(1-x)(1-y)}{y}\times\nonumber\\
\Big(\theta^2+\frac{x(1-x)}{y}\alpha^2+\frac{1}{2yx(1-x)}\frac{m^2}{E_a^2}\Big)^{-2}.
\end{align}

We first integrate over $\alpha$ and $\theta$ using Mathematica~\cite{mathematica}. We then expand in $m$ and carry out the rest of the integrals to find
\beq\nonumber
\frac{\lambda^2}{12(4\pi)^4}\log\frac{E_a^2\delta^2}{m^2}.
\eeq

For each leg, this correction exactly cancels the mass divergence that comes from $R$,
\beq\nonumber
R+\frac{\lambda^2}{12(4\pi)^4}\log\frac{E_a^2\delta^2}{m^2}=1+\frac{\lambda^2}{12(4\pi)^4}\Big(\log \delta^2 +\log\frac{E_a^2}{\mu^2}\Big).
\eeq

We note that in this theory, there are no double logs at the first order and just a single log for collinear divergence.

It is not as straight forward to find an expression for the AP function as it was in the cubic theory in six dimension (or as it is in QED and QCD). It is mainly because we cannot separate the integrals over the angles from those of the energies. The virtuality for particle $a$ in Fig. \ref{fig:phi4splitting} is
\begin{equation}\label{eq:virtualityphi4}
p_a^2=\frac{1}{z x (1-z-x)}\Big[ z(1-z) k_{T_c}^2+x(1-x)k_{T_d}^2-2 z x k_{T_c}k_{T_d} \cos \phi\Big].
\end{equation}
where $x=E_b/E_a$, $z=E_c/E_a$ and $\phi$ is the azimuthal angle between $c$ and $d$. In appendix \ref{ap:APphi4}, we have found that the AP function, in the massless theory, is
\begin{align}\nonumber
\mathcal{P}(x,z,t',t'')=\frac{\lambda^2}{3!(4\pi)^4}\frac{ x z (1-x-z)[x(1-x)t'+z(1-z)t'']}{\Big[(x(1-x)t'+z(1-z)t'')^2-4 x^2 z^2t't''\Big]^{3/2}},
\end{align}
where $t'=K_{T_b}^2$, and $t''=K_{T_c}^2$. Knowing the AP function, we find the Sudakov factor:
\begin{align}\nonumber
\Delta(t,t_0)=&\exp \Big[ -\int\PP(x,z,t,t')dzdxdtdt'\Big].
\end{align}

 We were not able to perform this integral analytically\footnote{We can evaluate this integral for parts of the integration interval: $z,d\in [0,1], t\in [\delta,1]$ and $t'<t$, using a combination of integration and expansion in $\delta$, and we still get the single logarithm of $\delta$, which is in agreement with equation \eqref{eq:APfunctionphi4}. }. Nor we could not find a simplified expression in terms of the virtuality \eqref{eq:virtualityphi4}.  But, from our other choice of variables in \eqref{eq:phi4splitting}, we know that the divergent part should be
\beq\label{eq:APfunctionphi4}
\int\PP(x,z,t',t'')dzdxdt'dt''\approx \frac{\lambda^2}{12(4\pi)^4}\log\frac{t}{t_0}.
\eeq
 where $t$ and $t_0$ are the hard and soft scales, and $t\gg t_0$. We also note that
\begin{itemize}
\item There are no soft divergences. The splitting function does not have a pole at $x,z=0,1$. We have checked this numerically. 
\item The integrals in \eqref{eq:APfunctionphi4} and \eqref{eq:phi4splitting} are only divergent if all the three final particles are collinear. If two are collinear and the third is not, the integral vanishes as $\delta\to 0$.
\end{itemize}
 Hence, we will use an effective AP function which is given by
 \beq\nonumber
 \PP(x,z,t)=\frac{\lambda^2(t)}{12(4\pi)^4}\frac{1}{t},
 \eeq
 where $t$ is the virtuality\footnote{The numerical pre-factor is likely different, but is irrelevant to our discussion here.}. We also have added the energy dependence of the coupling constant. 
 
The Sudakov factor, is given by
\begin{align}\label{eq:SFphi4}
\Delta(t,t_0)=&\exp \Big[ -\int\PP(x,z,t')dzdxdt'\Big]=\exp \Big[ -\int\frac{\lambda^2(t')}{12(4\pi)^4}\frac{dt'}{t'}\Big].
\end{align}
We use the two loop beta function to match our order of calculation of the AP function. We have for the quartic theory,
\begin{align}\nonumber
\beta(\lambda)&=3\frac{\lambda^2}{(4\pi)^2}-6\frac{\lambda^3}{(4\pi)^3},\\\nonumber
\frac{dt}{t}&=2\frac{d\lambda}{\beta(\lambda)}.
\end{align}
The Sudakov factor is then,
\begin{align}\nonumber
\Delta(t,t_0)&=\exp \Big[ \frac{1}{36(4\pi)^2}\log\frac{1-2\lambda(t)^2/(4\pi)^2}{1-2\lambda(t_0)^2/(4\pi)^2}\Big]\\
&=\Bigg[\frac{1-2\lambda(t)^2/(4\pi)^2}{1-2\lambda(t_0)^2/(4\pi)^2}\Bigg]^{b},\label{eq:Sudakovphi4}
\end{align}
where $b=\frac{1}{36(4\pi)^2}$.

In writing down the equation for the JGF \eqref{eq:jetGFgeneral}, we need to take into account that a line splits into 3 other lines. Using the facts presented in the previous section, the equation for the generating function reads:
\begin{align}\nonumber
\Phi_{(4)}[t]=&\Delta[t,t_0]\Phi_{(4)}[t_0]\\\nonumber
&+\int dx\int dz\int_{t_0}^t d t'
\Delta(t,t')\PP(x,z,t')\times\\\nonumber
&\quad\quad\Phi_{(4)}[x^2t']\Phi_{(4)}[z^2t']\Phi_{(4)}[(1-x-z)^2t'].
\end{align}

 Note that in this equation we use $t=p_a^2$, the virtuality \eqref{eq:virtualityphi4}, as the variable. As before, the first term on the right hand side takes care of the probability of not splitting from the soft scale $t_0$ to the hard scale $t$, and the second term takes into account when the particle splits at some intermediate scale $t'$. Since there is no divergence when only two lines are collinear, we have to add three GF in the second term. And, since there is no soft divergence, the integrand is not concentrated around $x,z\approx 0,1$. The energy dependences under the integral, in the generating functions and the coupling constant, are all logarithmic and we simplify by ignoring the energy dependence of these functions under the integral. 

Using $\Delta(t,t')=\Delta(t,t_0)/\Delta(t',t_0)$, and differentiating with respect to $t$ from both sides, we find
\beq\nonumber
\frac{d \Phi_{(4)}[t]}{dt}=\frac{1}{\Delta(t,t_0)}\frac{d\Delta(t,t_0)}{dt}\Big(\Phi_{(4)}[t]-\Phi_{(4)}^3[t]\Big).
\eeq

Solving this equation with the boundary condition $\Phi[t_0]=u$, we find
\beq\nonumber
\Phi_{(4)}[t]=\frac{u}{\sqrt{u^2+(1-u^2)\Delta^{-2}(t,t_0)}}.
\eeq

\begin{figure}[ht!]
\centering
\includegraphics[width=0.7\textwidth]{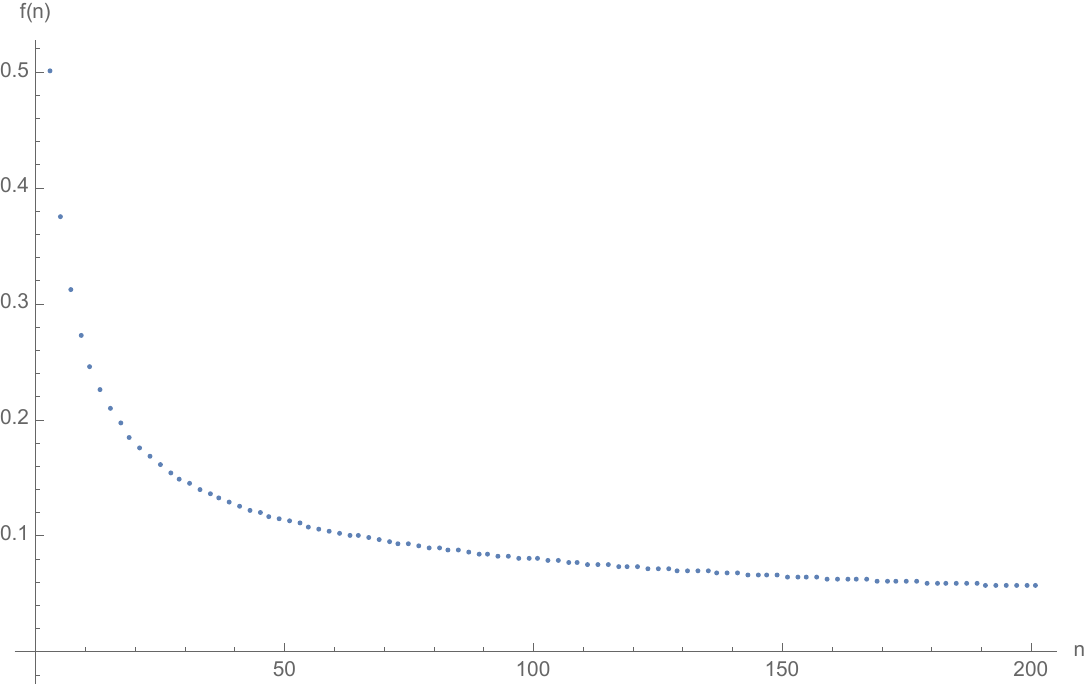}
\caption{$f(n)$ in equation \eqref{jetratephi4}.}
\label{fig:jetratefnPhi4}
\end{figure}

Using Mathematica~\cite{mathematica}, we find the jet rate \eqref{eq:jetrate},
\beq\label{jetratephi4}
R_n=f(n)\Delta\Big(1-\Delta^2\Big)^{\frac{n-1}{2}}\quad\quad n = \text{ odd}\geq 3,
\eeq
with $f(n)$ a slowly decreasing function of $n$, plotted in Fig.~\ref{fig:jetratefnPhi4}.

The average number of jets is,
\beq\nonumber
\braket{n}=\frac{d\phi(u,t)}{du}\Big|_{u=1}=\Delta^{-2}.
\eeq

where in both equations above, the Sudakov factor is given by \eqref{eq:Sudakovphi4}. These equations are again non-perturbative and have no factorial growth.

\section{\texorpdfstring{$\phi^3$}{phi32} Theory In Four Spacetime Dimensions}\label{sec:phi34D}

For completeness, let us briefly look at the cubic theory in four dimensions. The Lagrangian is
\begin{equation}\label{eq:phi34d}
\LL= \frac12\big(\partial\phi\big)^2-\frac12m^2\phi^2-\frac{g}{3!}\phi^3.
\end{equation}
 This is an example of a super-renormalized theory, and is not perturbativaly predictable in the IR. In the IR, one finds a series in $g^2/m^2$. It might seem that one cannot take the $m^2\to 0$ limit in this theory. However, we have found that for $Q^2 \gg m^2$, the IR divergences sum to a Sudakov form factor and the total cross section would be independent of the $m$. In other words, in this limit one finds a series in $g^2/Q^2$. Since in a general scalar theory, the cubic and the quartic couplings do not necessarily correlate, this cancelation is independent of quartic coupling. Hence, we can assume there exists a $\lambda\phi^4$ term with a negligible quartic coupling $\lambda\ll 0$ to avoid an unstable vacuum.
 

To show the IR divergences cancelation, it is possible to set $m=0$ and use dimensional regularization for both IR and UV divergences. We have done this in Appendix \ref{sec:phi34dIRcanc} for the process $\phi^*\to\phi\phi$, where $\phi^*$ is off-sehll. We found that the total cross section is,
\begin{align}\label{eq:totalXsecphi36d}
\sigma_{\phi^*\to 2\phi+3\phi}= \frac{g^2}{16\pi}\Bigg(1-\frac{3g^2}{16\pi^2 Q^2}\Bigg).
\end{align}

Keeping the mass and taking the limit $m\to 0$, the total cross section should be the same as \eqref{eq:totalXsecphi36d}. We skip a detail calculation and point out that the relevant virtual diagrams are the vertex corrections, Fig~\ref{fig:phi3vertexcorrection}, and the leg correction. In 4D, the vertex correction does indeed has a singularity (regardless of the scheme, that is the feynman parameter integral diverge if we set $m=0$). The divergence is a double log,
\begin{align}\nonumber
F_3(q^2)&=\frac{g^2}{(4\pi)^2}\int dxdydz\delta(x+y+z-1)\frac{1}{-xyq^2+(z^2-z+1)m^2},\\\label{eq:phi34dvertexcorrection}
&\approx \frac{g^2}{(4\pi)^2}\frac{1}{-q^2}\log^2\frac{-q^2}{m^2},
\end{align}
where in the last line we have used the tricks used for QED vertex correction to integrate\footnote{We have renamed one of the masses as the mass of would be ``photon" in a similar QED vertex, and integrate it as in QED \cite{Peskin:1995ev}. The result were checked numerically.}. This divergent, however, is not relevant. Since the coupling is dimension full, the contribution comes with a $g^2/Q^2$, which is suppressed in the limit we are working in. The other mass singularity comes from the residue of the mass pole. It is given by,
 \beq\nonumber
 R^{-1}=1+\Pi'(m^2),
 \eeq
 where as usual prime is differentiation with respect to $p^2$. We regulate the UV divergence with $d=4-\epsilon$ andfind,
 \beq\nonumber
 \Pi(p^2)=\frac{g^2}{2(4\pi)^2}\Gamma(\epsilon/2)\int_0^1 dx\frac{1}{(m^2-p^2x(1-x))^{\epsilon/2}}.
 \eeq
 We first differentiate with respect to $p^2$, and then expand in $\epsilon$, to find
 \beq\nonumber
 \Pi'(m^2)=\frac{g^2}{2(4\pi)^2}\int_0^1 dx\frac{x(1-x)}{m^2(1-x(1-x))}.
 \eeq
 The integral is finite and can be calculated using Mathematica~\cite{mathematica}. Hence, the residue becomes,
 \beq\label{eq:Rphi34d}
 R^{-1}=1-\frac{g^2}{(4\pi)^2}\frac{\frac12-\frac{\pi}{3\sqrt{3}}}{m^2}.
 \eeq
 
 The singularity is power divergent and suppresses the divergence coming from the vertex correction \eqref{eq:phi34dvertexcorrection}. 
 
 To find the radiation contribution we find the splitting shown in Fig.~\ref{fdiag:phi3s6dplit}. The contribution to the total cross section is
 \beq\nonumber
\frac{1}{2!}\int \frac{g^2}{(p_a^2-m^2)^2}d^4p_bd^4p_c.
\eeq
 
 Following the procedure outlined after \eqref{eq:split2},  we find the splitting probability when the angle between the two is less than $\delta$ to be
\beq\label{eq:serindikiphi3phi4}
\frac{g^2}{(4\pi)^2}\int_0^1 dz\int_0^{\delta}d\theta\frac{\theta}{Q^2z(1-z)\Big(\theta^2+m^2/Q^2f(z)\Big)^2},
\eeq 
where $f(z)=(1-z+z^2)/(z-z^2)^2$.

Integrating over the $\theta$, we find
\beq\label{eq:massivesplittingfunction1}
\frac{1}{2z(1-z)}\Bigg(\frac{1}{m^2 f(z)}-\frac{1}{Q^2\delta^2+m^2f(z)}\Bigg).
\eeq

If $\delta^2<m^2/Q^2$, the expression is finite as $\delta\to0$. In fact, this term vanishes for $\delta=0$, which means that these diagrams do not contribute to the jets with fewer legs.

However, if we take the limit $m^2/Q^2\to 0$ and $\delta\to 0$, while keeping $\delta^2>m^2/Q^2$, the above expression would become singular both in $m^2/Q^2$ and $\delta^2$. Integrating over $z$ first and then expanding \eqref{eq:massivesplittingfunction1} in $m^2/Q^2$, we find

\beq\label{eq:pertexpofAPforMphi34d}
\frac{g^2}{2(4\pi)^2}\Big[\Big(-1+\frac{2\pi}{3\sqrt{3}}\Big)\frac{1}{m^2}+\frac{1}{Q^2\delta^2}+\mathcal{O}(m^2)\Big].
\eeq

As expected, this expression cancels the $m^2$ divergence of $R$ \eqref{eq:Rphi34d}\footnote{From LSZ formula, each leg is multiplied by $\sqrt{R}$ in the S matrix, and by $R$ in the cross section.}.

We can use \eqref{eq:serindikiphi3phi4} to find AP function. However, we should be careful, since we cannot set $m=0$ in \eqref{eq:serindikiphi3phi4} without changing the variables. It gives a wrong result since $f(z)$ has poles at $z=0,1$. We have to first change the variable to $k_T$. The integrand becomes, (see appendix~\ref{atphi34d} for another derivation)
 
 \begin{align}\nonumber
\PP(z,t) &=\frac{g^2}{(4\pi)^2}\frac{z(1-z)}{\Big(t+ m^2 (1-z+z^2)\Big)^2},
\end{align}
 
 where $t=k_T^2$. Now, setting $m=0$ we find
 \begin{align}\label{splittingfucntion}
\PP &=\frac{g^2}{(4\pi)^2}\frac{z(1-z)}{t^2}.
\end{align}

 We have a power divergence, called \emph{ultra-collinear} divergence, instead of a logarithmic one\footnote{This is the same splitting function as found in~\cite{Chen:2016wkt}. See Table 2 of that paper.}.

In four dimensions, the renormalized coupling is the same as the bare one, since the vertex diagram \ref{fig:phi3vertexcorrection} is not divergent. Hence, the Sudakov form factor can be find by integrating \eqref{splittingfucntion}:
\begin{align}
\Delta(t,t_0)&=\exp \bigg[ - \int_{t_0}^t dt' \int dz \PP(z,t')\bigg]\nonumber,\\
&\approx\exp \bigg[ - \frac{g^2}{6(4\pi)^2 t_0}.\bigg].\label{SF4dphi3}
\end{align}

 The equation for the generating function does not change with the dimension. The equation \eqref{jgf6d} for the cubic theory in six dimensions is good here as well. We again find \eqref{phi36dGF}:

 \beq\nonumber
\Phi_{(3)}[t]=\frac{u}{u+(1-u)\Delta^{-1}},
\eeq

with jet rates given by \eqref{eq:Rnph36d},
\beq\nonumber
R_n=\Delta(1-\Delta)^{n-1}.
\eeq

\section{Off-Shell \texorpdfstring{$\phi*$ $\to n\phi$}{phitonphi} Process. }\label{sec:offsh}

 The generating functions found in the previous sections, correspond to the evolution of a highly relativistic and approximately on-shell particle. Hence, to apply this method to the process $\phi^*\to n\phi$, where the $\phi^*$ is highly off-shell, we approximate the JGF for this process as
\beq\nonumber
\Phi_{\phi^*\to n\phi}[t]= \Phi_{(3)}[t/4]^2+\dots,
\eeq
for the cubic theory and
\beq\nonumber
\Phi_{\phi^*\to n\phi}[t]= \Phi_{(4)}[t/9]^3+\dots,
\eeq
for the quartic theory. The next terms will be of order $\OO(\lambda^2)$ with additional Sudakov factors, and highly suppressed.

\begin{figure}[b!]
\centering
\includegraphics[width=0.7\textwidth]{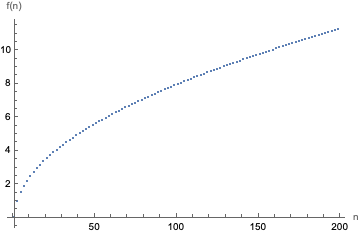}
\caption{$f(n)$ function in equation \eqref{eq:Rnph36d2-2}}
\label{fig:jetratefnPhi4offshell}
\end{figure}

For the cubic and quartic theories, the jet rate becomes
\beq 
R^{(3)}_n=(n-1)\Delta^2(1-\Delta)^{n-2}, \label{eq:Rnph36d2-1}\\
R^{(4)}_n=f(n)\Delta^3(1-\Delta^2)^{\frac{n-3}{2}}, \label{eq:Rnph36d2-2}
\eeq
with $f(n)$ given in Fig.~\ref{fig:jetratefnPhi4offshell}. These expressions do not change with spacetime dimensions.

\begin{figure}[ht!]
\centering
\begin{subfigure}{0.49\textwidth}
\includegraphics[width=\textwidth]{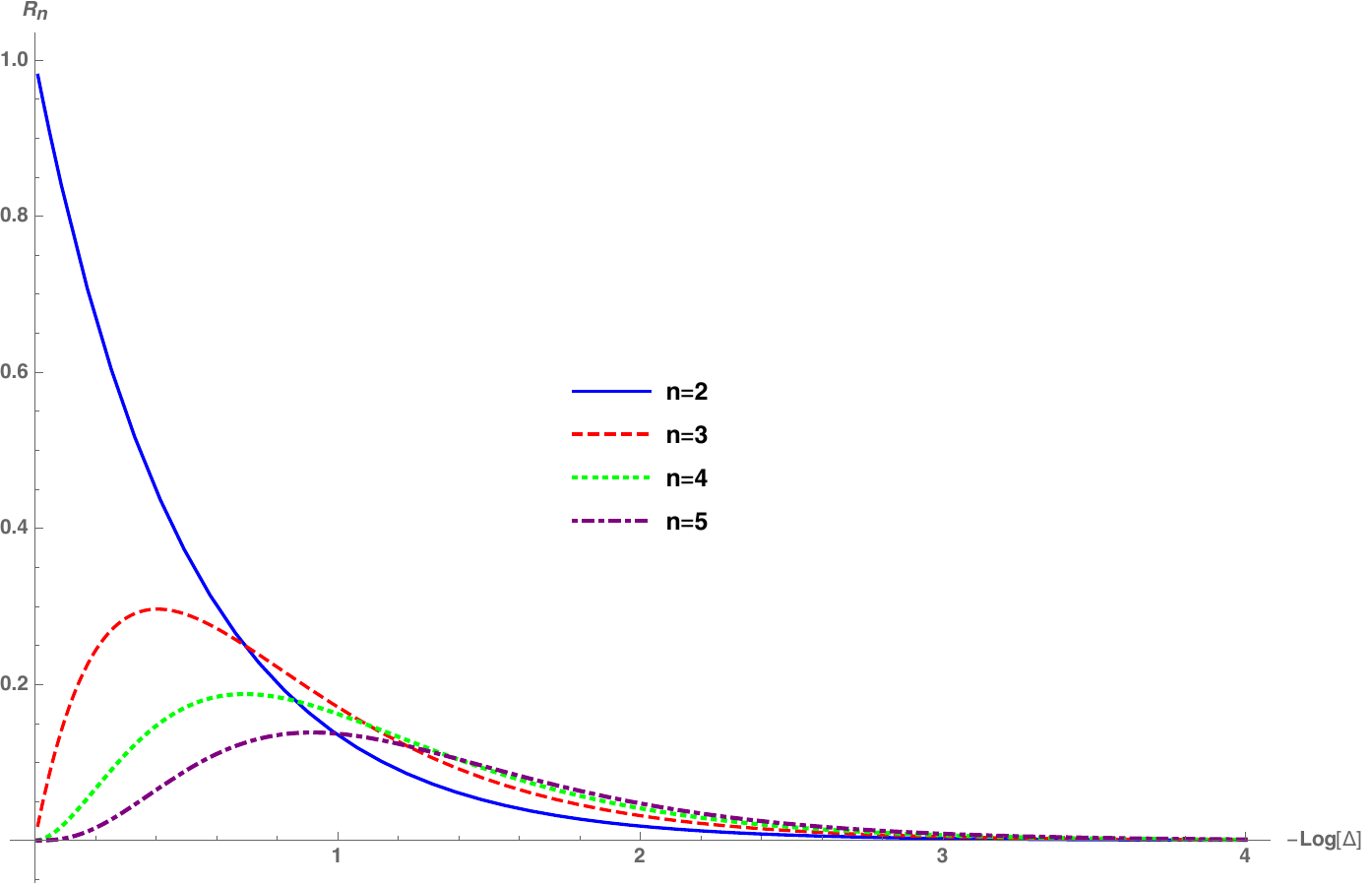}
\caption{The cubic theory.}
\label{fig:jetratePhi3and4-1}
\end{subfigure}
\begin{subfigure}{0.49\textwidth}
\includegraphics[width=\textwidth]{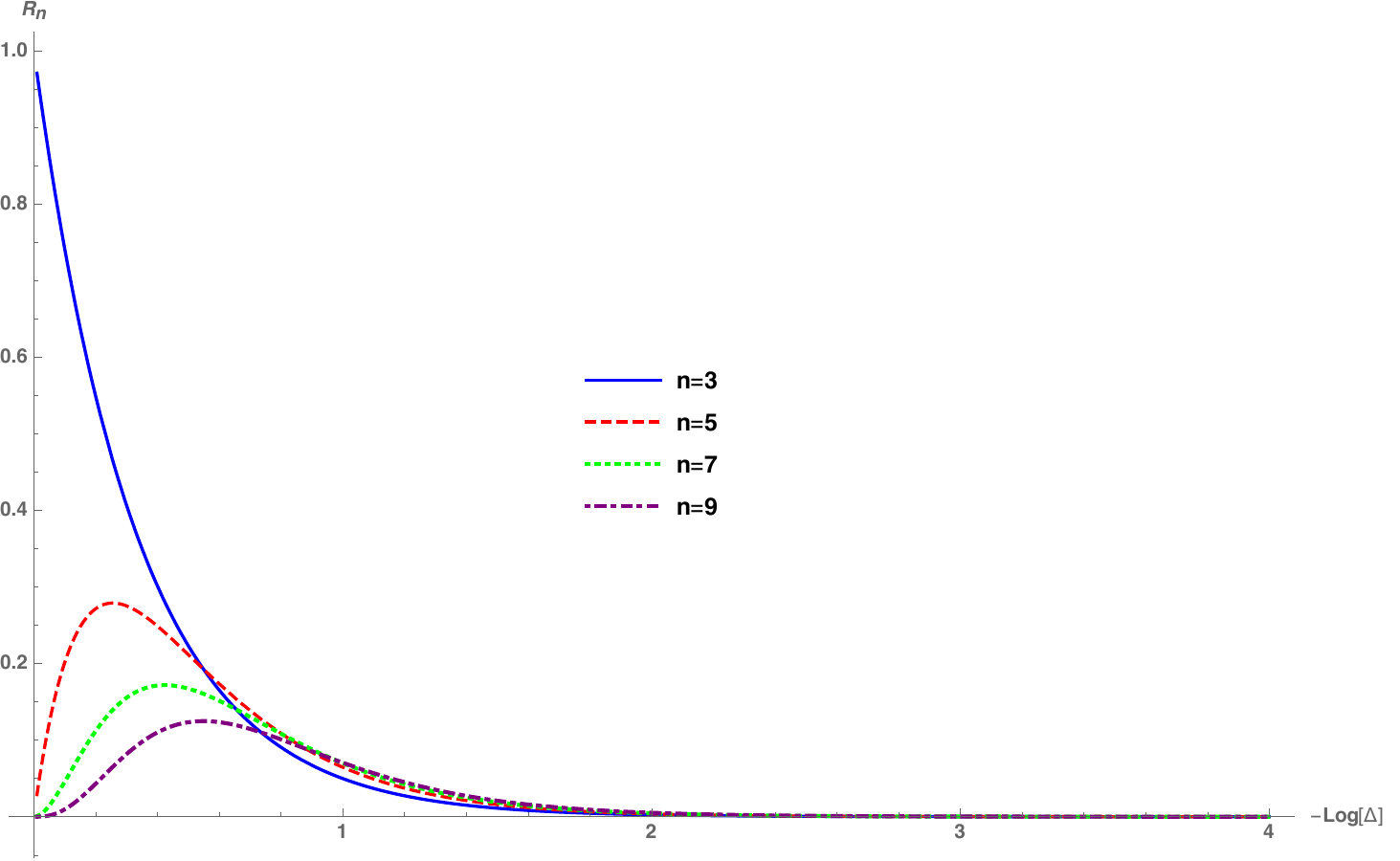}
\caption{The quartic theory.}
\label{fig:jetratePhi3and4-2}
\end{subfigure}
\caption{The jet rates.}
\end{figure}

In Fig.~\ref{fig:jetratePhi3and4-1} and~\ref{fig:jetratePhi3and4-2}, we have plotted few of the jet rates as a function of $-\log[\Delta]$. We can see that the multi-scalar rates start to dominate when
\beq\nonumber
-\log[\Delta]>1.
\eeq
For $\phi_{\text{6D}}^3$, this corresponds to
\beq\nonumber
\frac{t}{t_0}>\exp \big[\frac{12(4\pi)^3}{g^2}\big].
\eeq
This is an extremely large number for any perturbative value of the coupling constant. Hence, almost always we will see two or three jets in any high energy process involving only scalar particles. Similar situation holds for the other theories.

This also means that we can expand the Sudakov factor in the powers of the coupling constant while still keeping the non-relativistic limit, since for most $t>t_0$ we have $\Delta\approx 1$.  Using \eqref{eq:Sudakovfactorphi36d}, \eqref{eq:Sudakovphi4}, and \eqref{SF4dphi3}, we find the expressions 
\begin{align}\label{eq:ratephi36dexpandoffshell-0}
R_n^{\phi^3_{\text{6D}}}=&(n-1)\Big(\frac{g_6^2}{12(4\pi)^3}\log \frac{t}{t_0}\Big)^{n-2}\Big(1-\frac{g_6^2}{6(4\pi)^3}\log \frac{t}{t_0}\Big),\\\label{eq:ratephi44dexpandoffshell-0}
R_n^{\phi^4_{\text{4D}}}=&f(n)\Big(\frac{\lambda^2}{6(4\pi)^4}\log \frac{t}{t_0}\Big)^{\frac{n-3}{2}}\Big(1-\frac{\lambda^2}{4(4\pi)^4}\log \frac{t}{t_0}\Big),\\\label{eq:ratephi34dexpandoffshell-0}
R_n^{\phi^3_{\text{4D}}}=&(n-1)\Big(\frac{g_4^2}{6(4\pi)^2t_0}\Big)^{n-2}\Big(1-\frac{g_4^2}{3(4\pi)^2t_0}\Big),
\end{align}
for the particle rates in $\phi^3_{\text{6D}}$, $\phi^4_{\text{4D}}$, and $\phi^3_{\text{4D}}$ theories respectively. For the $\phi^3_{\text{4D}}$, when $t>t_0$, we have found that the Sudakov factor doesn't depend on $t$, and the expansion holds when $g_4^2<t_0$.

Following our discussion after the equation \eqref{eq:twototwocrosssectionphi36d}, let us make an ansatz on the particle cross section using the jet rates above. There, we discussed that we can use the mass as the cutoff $t_0$. In this case, the Sudakov factor just sums the virtual contributions as can be shown in $\phi^3_{\text{6D}}$ using SCET \cite{Becher:2014oda}. Plugging $t=Q^2$ and $t_0=m^2$ in  \eqref{eq:ratephi36dexpandoffshell-0}, \eqref{eq:ratephi44dexpandoffshell-0}, and \eqref{eq:ratephi34dexpandoffshell-0}, we find

\begin{align}\label{eq:ratephi36dexpandoffshell}
R_n^{\phi^3_{\text{6D}}}=&(n-1)\Big(\frac{g_6^2}{12(4\pi)^3}\log \frac{Q^2}{m^2}\Big)^{n-2}\Big(1-\frac{g_6^2}{6(4\pi)^3}\log \frac{Q^2}{m^2}\Big),\\\label{eq:ratephi44dexpandoffshell}
R_n^{\phi^4_{\text{4D}}}=&f(n)\Big(\frac{\lambda^2}{6(4\pi)^4}\log \frac{Q^2}{m^2}\Big)^{\frac{n-3}{2}}\Big(1-\frac{\lambda^2}{4(4\pi)^4}\log \frac{Q^2}{m^2}\Big),\\
R_n^{\phi^3_{\text{4D}}}=&(n-1)\Big(\frac{g_4^2}{6(4\pi)^2m^2}\Big)^{n-2}\Big(1-\frac{g_4^2}{3(4\pi)^2m^2}\Big).
\end{align}

In general, the numerical pre-factors of the divergent terms - that come from the Sudakov factors: $\frac{1}{12}$, $\frac{1}{12}$ (which is $\frac{1}{6}$ in the equation above since Sudakov factor is squared in \eqref{eq:Rnph36d2-2}), and $\frac{1}{6}$ for $\phi^3_{\text{6D}}$, $\phi^4_{\text{4D}}$, and $\phi^3_{\text{4D}}$ respectively -  change with changing the cut-off parameter, $t$. Hence, these factors in the above expressions might not be correct when we changed the cut-off parameter to the mass. Without derivation, by matching to perturbative result, we can guess the correct pre-factors. Since we have found that in $\bar{\text{MS}}$ scheme, the mass divergence comes from the residue at the mass pole, $R$, we can see from \eqref{eq:Rphi36d} for $\phi^3_{\text{6D}}$ the numerical factor is in fact $\frac{1}{12}$. In section \ref{sec:phi44D}, we guessed an effective splitting function for $\phi^4_{\text{4D}}$ based on the perturbative result which lead to \eqref{eq:Rphi44d}, hence $\frac{1}{12}$ is automatically correct. For $\phi^3_{\text{4D}}$, from \eqref{eq:Rphi34d}, we can see that the correct pre factor is $c=(\frac{1}{2}-\frac{\pi}{3\sqrt{3}})$. Hence, we find,
\beq\label{eq:ratephi34dexpandoffshell}
R_n^{\phi^3_{\text{4D}}}=&(n-1)\Big(c\frac{g_4^2}{(4\pi)^2m^2}\Big)^{n-2}\Big(1-2c\frac{g_4^2}{(4\pi)^2m^2}\Big).
\eeq

\section{Discussion and Conclusion}\label{sec:conclusion}
  We have shown that the IR divergences cancel in scalars theories. We found that there is only a single logarithm divergence in these theories, corresponding to collinear divergent. Hence, we defined scalar jets with the opening angle $\delta$ and found the Altarelli-Parisi functions\footnote{Contrary to what is usual, we have moved the $t$ dependence into the definition of $\mathcal{P}$ since in these cases it does not always factor out and can have different powers.}. They are given in \eqref{eq:APfunctionPhi36d}, \eqref{eq:APfunctionphi4}, and \eqref{splittingfucntion}, for $\phi^3_{\text{6D}}$, $\phi^4_{\text{4D}}$, and $\phi^3_{\text{4D}}$ respectively. 

The Sudakov factor, by definition, is the probability of no-splitting. They sum the IR divergent terms. They are given in \eqref{eq:Sudakovfactorphi36d}, \eqref{eq:Sudakovphi4}, and \eqref{SF4dphi3}, for $\phi^3_{\text{6D}}$, $\phi^4_{\text{4D}}$, and $\phi^3_{\text{4D}}$ respectively. 

We used the Sudakov factor to write an equation for the generating function of jet rates. The jet rates, for $\phi^*\to n\phi$ processes, are given by equations \eqref{eq:Rnph36d2-1} and \eqref{eq:Rnph36d2-2}, for the cubic theory and the quartic theory respectively. These formulas do not depend on the spacetime dimensions. The jet cross sections are given by $\sigma_n=R_n\sigma_{\text{total}}$, where the total cross section corresponds to fix $Q$ (note that $\sum_n R_n=1$). For the broken theory, in the limit $Q^2\gg v^2$, the generating function is approximately the same as the quartic theory generating function.

We argued that for the massive theory, if $\delta$ is smaller than $m/Q$, where Q is the hard scale, we can substitute $\delta$ with $m/Q$ in the Altarelli-Parisi function and consequently in the Sudakov factor. Expanding the rates in power of the couplings we found \eqref{eq:ratephi36dexpandoffshell}, \eqref{eq:ratephi44dexpandoffshell}, and \eqref{eq:ratephi34dexpandoffshell}, for $\phi^3_{\text{6D}}$, $\phi^4_{\text{4D}}$, and $\phi^3_{\text{4D}}$ respectively.  In appendix \ref{sec:fixorder}, we have checked the 4D theories in \emph{Madgraph}. Due to large number of diagrams, it is not possible to compute the cross sections for more than few particles in the final state. For the few cross sections we calculated, our result matches the numerical calculation in the limit $m^2\to 0$. 

 To summaries, if we set
\beq\label{eq:epsilondef}
Q=nm(1+\epsilon),
\eeq
where $\epsilon$ is the average kinetic energy (in units of $m$) of the final particles, the jet calculation is valid in the limits $\epsilon\gg1$. While the threshold limit, which was the center of the most previous analysis, is at $\epsilon \ll 1$. Nevertheless, we can state few things regarding multi-scalar cross sections before we finish:

 \begin{itemize}
 \item Our result does not support the Holy-Grail function. There is no dependence on $n\lambda$, but on $n$ and $\lambda$.
 
 For the sake of interest, let us try to write our result for the unbroken $\phi^4_{\text{4D}}$ in the form of the Holy-Grail function~\eqref{hollygrail}. Since the exponent of the Sudakov factor is small even for very large $n$, we use the expansion in~\eqref{eq:ratephi44dexpandoffshell}. Approximating the function $f(n)\approx \sqrt {n/1.6}$ for large $n$,  we find
\begin{align}\nonumber
\sigma_n^{\phi^4_{\text{4D}}}&=R_n^{\phi^4_{\text{4D}}}\sigma_{\text{total}}\\\label{crosssectionphi44d}
&\approx \exp[\frac{1}{2}\log n  +n \log \lambda +n\log\log n (1+\epsilon)  + \text{const} ]\sigma_{\text{total}}.
\end{align}

Hence, the Holy-Grail function is
\begin{align}\nonumber
F=\frac{1}{2}\lambda \log n  +\lambda n \log \lambda +\lambda n\log\log n (1+\epsilon)  + \lambda \times \text{const} .
\end{align}

However, we should be careful since the total cross section in~\eqref{crosssectionphi44d} depends on $Q^2$, and since we used \eqref{eq:epsilondef}, it depends on $n$. Perhaps a better quantity for comparison would be $\sigma_{n+1}/\sigma_{n}$ at fixed $Q^2$ \footnote{If one tries to calculate this quantity, they should note that for fixed $Q^2$, equation \eqref{eq:epsilondef} will give different average energies, $\epsilon$, for different number of final particles.}.

 \item Our result is in contradiction with the Higgsplosion proposal which states that the cross section will become unsuppressed at high energy. Note that the region where the final Higgses becomes unsuppressed, cover the relativist ($\epsilon\approx 10$) region (see section \ref{sec:multihiggs}). A more recent analysis based on the same method as in Higgsplosion analysis\cite{Demidov:2022ljh}, is extended to relativistic regime. Though their result \eqref{eq:Demidov2022} differs with ours, it does not support a factorial growth neither. 
 
 \item A jet contains all the particles that can be fitted into the jet cone and jet energy. The fact that even the two jet cross section is finite as $Q\to\infty$, indicates that the factorial divergence does not exist. 
 \item We have found that while the final particles are relativistic, the cross sections do not grow factorially. It is thus hard to believe that the dependence of the cross section on the number of the final particles change as we change the energy of the particles. It is interesting to see why the semi-classical methods sometimes give the factorial divergent. It is also interesting to see why in the calculations that sum the leading virtual corrections to all order, the $\log m/Q$ does not show up. 
 \end{itemize}

ACKNOWLEDGMENT: This work was the continuation of part of my PhD thesis, done in the University of California in Davis. Afterwards, parts of it was completed during my postdoc position at Sharif University of Technology. Without Dr. John Terning guidance and help this work would not have been possible. I thank Miranda Chen for sharing her calculation with me. I thank Fayez Abu-Ajamieh, with whom the early stages of this work was developed. I am also thankful to Dr. Mahdi Torabian for hosting me in the Sharif University.

\newpage
\begin{appendices}
\section{Loop Corrections to All Order at Threshold for \texorpdfstring{$\phi^3_{\text{6D}}$}{phi36d}}\label{app:loopPhi36d}
The one-loop correction to multi-scalar process in the cubic theory can be find using the methods in~\cite{Argyres:1993wz, Papadopoulos:1993aw, Smith:1992rq}.

Let us define $a_i(n)$ as the amplitude of producing $n$ particle at $i$'th loop order.
We have
\begin{align}\nonumber
\phi_0(x)=&\sum\frac{ia_0(n)x^n}{n!(n^2-1)},\\
\phi_1(x)=&\sum\frac{ia_1(n)x^n}{n!(n^2-1)}.\label{eq:phi1phi36D}
\end{align} 

 \begin{figure}[ht!]
   \begin{center}
   \begin{eqnarray*} 
   \parbox{20mm}{ 
\begin{fmffile}{loopPhi36d_1}
    \begin{fmfgraph}(60, 30)
     \fmfleft{l1}
     \fmfright{r1}
     \fmf{dashes}{l1,o}
     \fmf{phantom}{o,r1} 
               \fmfv{d.sh=circle,d.f=empty,b=(1,,0,,1),d.si=30,label=$n$}{o}
    \end{fmfgraph}
    \end{fmffile}
  }=&\quad
   \parbox{30mm}{ 
\begin{fmffile}{loopPhi36d_2}
    \begin{fmfgraph}(60, 30)
     \fmfleft{l1}
     \fmfright{r1,r2}
     \fmf{dashes,tension=3}{l1,o}
     \fmf{dashes}{o,r1}
     \fmf{dashes}{o,r2}
               \fmfv{d.sh=circle,d.f=empty,d.si=30,label=$n1$}{r1}
               \fmfv{d.sh=circle,d.f=empty,b=(1,,0,,1),d.si=30,label=$n2$}{r2}
    \end{fmfgraph}
    \end{fmffile}
  }+\quad
   \parbox{30mm}{ 
\begin{fmffile}{loopPhi36d_3}
    \begin{fmfgraph}(65, 35)
     \fmfleft{l1}
     \fmfright{r1}
     \fmf{dashes,tension=2}{l1,o}
     \fmf{phantom}{o,r1} 
               \fmfv{d.sh=circle,d.f=empty,d.si=30,label=$n$}{r1}
          \fmffreeze
	\fmf{dashes,right,tension=0}{o,r1}
	\fmf{dashes,left,tension=0}{o,r1}
    \end{fmfgraph}
    \end{fmffile}
  }\nonumber\\ \nonumber\\
  &+\quad
  \parbox{30mm}{ 
\begin{fmffile}{loopPhi36d_4}
    \begin{fmfgraph}(75, 35)
     \fmfleft{l1}
     \fmfright{r1}
     \fmf{dashes,tension=2}{l1,o}
     \fmf{dashes}{o,o1}
     \fmf{phantom}{o1,r1} 
               \fmfv{d.sh=pentagram,d.f=full,d.si=8,label=$n$}{o}
               \fmfv{d.sh=circle,d.f=empty,d.si=30,label=$n$}{o1}
    \end{fmfgraph}
    \end{fmffile}
  }+\quad
  \parbox{30mm}{ 
\begin{fmffile}{loopPhi36d_5}
    \begin{fmfgraph}(75, 35)
    \fmfleft{l1}
     \fmfright{r1,r2}
     \fmf{dashes,tension=3}{l1,o}
     \fmf{dashes}{o,r1}
     \fmf{dashes}{o,r2}
               \fmfv{d.sh=circle,d.f=empty,d.si=30,label=$n1$}{r1}
               \fmfv{d.sh=circle,d.f=empty,d.si=30,label=$n2$}{r2}
 \fmfv{d.sh=pentagram,d.f=full,d.si=8,label=$n$}{o}
    \end{fmfgraph}
    \end{fmffile}
  }
      \end{eqnarray*}    
      \end{center}
    \caption{The recursion relation for the one loop corrections.}
    \label{fig:loopPhi36d}
  \end{figure}

 We start by writing the one loop amplitude recursively via Fig.~\ref{fig:loopPhi36d}. We have
\begin{align}
\frac{a_1(n)}{n!}=&-ig\sum_{n_1,n_2}\frac{ia_1(n_1)}{n_1!(n_1^2-1)}\frac{ia_0(n_2)}{n_2^2(n_2^2-1)}\nonumber\\
&-i\frac{g}{2}\int\frac{d^Dk}{(2\pi)^D}\frac{D(n,k)}{n!} \nonumber\\
&-iT_2\frac{ia_0(n)}{n!(n^2-1)}\nonumber\\
&-iT_3\frac{ia_0(n_1)}{n_1!(n_1^2-1)}\frac{ia_0(n_2)}{n_2!(n_2^2-1)},\label{eq:phi36d1loopeq}
\end{align}
where $q=(1,0,0,0)$ and we have set the mass equal to unity in the denominators. The recursion relation for the propagator function, $D(n,k)$, is given in Fig.~\ref{fig:prpLoopPhi36d} and is
\begin{align}\label{eq:propequationphi36d}
\frac{D(n,k)}{n!}=&-ig\frac{i}{(k+nq)^2-1+i\epsilon}\sum\frac{ia(n_2)}{n_2!(n_2^2-1)}\frac{D(n_1,k)}{n_1!}.
\end{align}

\begin{figure}[ht!]
   \begin{center}
   \begin{eqnarray*} 
   \parbox{30mm}{ 
\begin{fmffile}{proploopPhi36d_1}
    \begin{fmfgraph}(65, 35)
     \fmfleft{l1}
     \fmfright{r1}
     \fmf{dashes}{l1,o}
     \fmf{dashes}{o,r1} 
               \fmfv{d.sh=circle,d.f=empty,d.si=30,label=$n$}{o}
          \fmffreeze
    \end{fmfgraph}
    \end{fmffile}
  }=\quad
   \parbox{70mm}{ 
\begin{fmffile}{proploopPhi36d_2}
    \begin{fmfgraph}(115, 85)
     \fmfleft{l1}
     \fmfright{r1}
     \fmftop{t1}
     \fmf{dashes}{l1,ol}
     \fmf{dashes}{ol,oc}
     \fmf{dashes}{oc,or}
     \fmf{dashes}{or,r1}
     \fmffreeze
      \fmfv{d.sh=circle,d.f=empty,d.si=30,label=$n1$}{or}
          \fmf{dashes}{oc,t1}
      \fmfv{d.sh=circle,d.f=empty,d.si=30,label=$n2$}{t1}
    \end{fmfgraph}
    \end{fmffile}
  }
      \end{eqnarray*}    
      \end{center}
    \caption{The recursion relation for the propagator.}
    \label{fig:prpLoopPhi36d}
  \end{figure}

 $T_2$ is the sum of mass and field counter-terms and $T_3$ is the vertex counter-term:
\begin{align*}
&-iT_2=-i\Big(m^2\delta_{m}+n^2m^2\delta_Z\Big),\\
&-iT_3=-i\delta_{g}.
\end{align*}
 The coefficient of the $\delta_Z$ term is $p^2=(nm)^2$ at threshold. From renormalization we know that
 \begin{align}\nonumber
 \delta_m=&-\frac{1}{\epsilon}\frac{g^2}{(4\pi)^2},\\\nonumber
 \delta_Z=&\frac{1}{6\epsilon}\frac{g^2}{(4\pi)^2},\\\nonumber
 \delta_{g}=&-\frac{1}{\epsilon}\frac{g^3}{(4\pi)^3}.
 \end{align}

 Let us also define,
\begin{align}\nonumber
f(x,k)=&\sum\frac{-iD(n,k)x^n}{n!}.
\end{align}
 Multiplying \eqref{eq:phi36d1loopeq} by $x^n$ and summing over $x$, we find an equation for $\phi_1$~\eqref{eq:phi1phi36D}:
 \begin{align}
 \Big(x\frac{d}{dx}\big(x\frac{d}{dx}\big)-1\Big)\phi_1(x)=&g\phi_1(x)\phi_0(x)+\frac g2\int\frac{d^Dk}{(2\pi)^D}f(x,k)\label{eq:phi36dloopeq1}\\
 &+m^2\Big[\delta_m\phi_0(x)+\delta_Z\Big(\phi_0(x)+\frac{g}{2}\phi_0^2\Big)\Big]\label{eq:phi36dloopeq2}\\
 &+\frac g2 \delta_{g}\phi_0^2.
 \end{align}

Writing $x=-\frac{12}{g}e^{\tau}$ and $f(x,K)=ye^{-\epsilon \tau}$ and $\epsilon=kq=k_0$ and $\omega=\sqrt{\vec{k}^2+1-i\epsilon}$, from \eqref{eq:propequationphi36d} the equation for $y$ would be
\begin{equation}\nonumber
\Big(\frac{d^2}{d\tau^2}-\omega^2+\frac{3}{\cosh \tau/2}\Big)y=e^{\epsilon\tau}.
\end{equation}

With $u=e^{\tau}$, the solutions are
\begin{align*}
f_1=&\big(u^{-\omega}(1+u)^{-3}\big)\big(3-27 u + 27 u^2 - 3 u^3 - 11 \omega + 27 u \omega + 
 27 u^2 \omega\\
 &- 11 u^3 \omega + 12 \omega^2 + 
 12 u \omega^2 - 12 u^2 \omega^2 - 12 u^3 \omega^2\\
 & - 4 \omega^3 - 12 u \omega^3 - 12 u^2 \omega^3 - 
 4 u^3 \omega^3\big),\\
f_2=&f_1(\omega\to-\omega).
\end{align*}
The Wronskian is $W=2 \omega (9 - 49 \omega^2 + 56 \omega^4 - 16 \omega^6)$. The general solution for $f$ is
\beq\nonumber
f(x,K)=\frac{e^{-\epsilon\tau}}{W}\Big(f_1\int d s e^{\epsilon s}f_2+f_2\int d s e^{\epsilon s}f_2\Big).
\eeq
The $K_0$ integral in \eqref{eq:phi36dloopeq1} gives a delta function and we find
\beq\nonumber
\frac g2\int\frac{d^Dk}{(2\pi)^D}f(x,k)=\frac g2\int\frac{d^dK}{(2\pi)^d}\frac{f_1f_2}{W},
\eeq
where $d=D-1$. It is possible to write this expression entirely in terms of $\phi_0$. We have
\begin{align}\nonumber
\frac g2\int\frac{d^dK}{(2\pi)^d}\Big[\frac{1}{2\sqrt{1+\vec{K}^2}}\Big]\Big[&\frac 12-\frac{g\phi_0}{3+4\vec{K}^2}+\frac{5g^2\phi_0^2}{(24+32\vec{K}^2)\vec{K}^2}\\\nonumber
&+\frac{25g^3\phi_0^2}{(15+8\vec{K}^2-16\vec{K}^4)24\vec{K}^2}\Big].
\end{align}

The first integral gives exactly the tadpole contribution. The divergent parts of the second and third terms are
\begin{align}\nonumber
&\frac g2\int\frac{d^dK}{(2\pi)^d}\frac{1}{2\sqrt{1+\vec{K}^2}}\frac{-g^2\phi_0}{6+8\vec{K}^2}=\frac{g^2}{(4\pi)^3}\frac{5}{6\epsilon}+const.\\\nonumber
&\frac g2\int\frac{d^dK}{(2\pi)^d}\frac{1}{2\sqrt{1+\vec{K}^2}}\frac{g^3\phi_0^2}{(24+32\vec{K}^2)\vec{K}^2}=\frac{g^3}{(4\pi)^3}\frac{5}{12\epsilon}+const.,
\end{align} 

where $d=6-\epsilon$ and $A$ and $B$ are some numerical constant. The divergence of the first integral is canceled by the $\OO(\phi_0)$ part of the counter-terms in \eqref{eq:phi36dloopeq2}. And the divergence of the second integral is canceled by the remaining part of $\delta_Z$ and $\delta_{g}$. 

The equation for $\phi_1$ boils down to

 \begin{align}\nonumber
 \Big(x\frac{d}{dx}\big(x\frac{d}{dx}\big)-1-g\phi_0\Big)\phi_1(x)=Ag^2\phi_0(x)+Bg^3\phi_0(x)^2_Cg^4\phi_0^3.
\end{align}
It is customary to set the finite part of the counter-terms such that $A$ and $B$ are zero~\cite{Argyres:1992np, Smith:1992rq}\footnote{If we don't do this, the solution has a log term, $\phi_1=\log x+\dots$, that is singular at $x=0$}.

With $A=B=0$, we find that
\beq\nonumber
\phi_1(x)=\frac{18g^2x}{(1-\frac{g}{12}x)^4}.
\eeq
We finally find the amplitude to the first order
\beq\nonumber
A_n=\frac{d^n}{dx^n}(\phi_0+\phi_1)\big|_{x=0}=nn!\Big(\frac{g}{12}\Big)^{n-1}\Big[1+3g^2(n^2+3n+2) \Big].
\eeq

\section{AP Function for \texorpdfstring{$\phi^4_{\text{4D}}$}{phi44d}}\label{ap:APphi4}
 When a particle splits to three particles, we can write the cross section with two less particles as
\begin{align}\nonumber
\sigma_{1\to n+c+d}&=\text{flux factor} \times \int d \Pi^f d \Pi^c d \Pi^d \quad | M_{n+c+d}|^2, \\\nonumber
& = \text{flux factor} \times \int d \Pi^f dz dx dk_{T_c}^2dk_{T_d}^2 \mathcal{P}_{\phi^4}(z,x,k_{T_c},k_{T_d}) |M_n|^2, \\\nonumber
&= \sigma_{1\to n} \int dz dx dk_{T_c}^2dk_{T_d}^2 \mathcal{P}_{\phi^4}(z,x,k_{T_c},k_{T_d}),
\end{align}
where $z=E_c/E_a$ and $x=E_d/E_a$. The Sudakov decomposition of momentum~\cite{Ellis:1991qj} is given by
\begin{align}\nonumber
p_b &=(1-z-x)p_a+(\beta+\alpha) u +K_{T_c}-K_{T_d}, \\\nonumber
p_c &= z p_a -\alpha u-K_{T_c},\\\nonumber
p_d &=xp_a-\beta u +K_{T_d} ,
\end{align}
so that $p_a=p_b+p_c+p_d$. The vector $u$ is chosen such that it is perpendicular to $K_{T_{c,d}}$ and $u^2=0$. Choosing it to be (1,0,0,-1), gives $u.p_a\approx 2 E_a$, assuming that $p_a$ is highly boosted. The phase-space integrals become
\begin{align}\nonumber
d\Pi^bd\Pi^c d\Pi^d &=d\Pi^a\frac{dz dz' d k_{T_c}^2dk_{T_d}^2 d\phi_cd\phi_d}{16(2\pi)^6 E_a^2z x(1-z-x)}.
\end{align}

 We can find $\beta$ and $\alpha$ by imposing on-shell condition for particles $c$ and $d$. Assuming that all the particles are massless, these conditions give
\begin{align}\nonumber
2\alpha n.p_a&=zp_a^2-K_{T_c}^2/z,\\\nonumber
2\beta n.p_a&=xp_a^2-K_{T_d}^2/x.
\end{align}

Using these equations and on-shellness of the final particles, we arrive at
\begin{equation}\nonumber
p_a^2=\frac{1}{z x (1-z-x)}\Big[ z(1-z) k_{T_c}^2+x(1-x)k_{T_d}^2-2 z x k_{T_c}k_{T_d} \cos \phi\Big],
\end{equation}
where $\phi$ is the angel between $K_{T_c}$ and $K_{T_d}$. Integrating over the azimuthal angels, we find the splitting function to be
\begin{equation}\nonumber
\mathcal{P}_{\phi^4}(z,x,k_{T_c},k_{T_d})=\frac{\lambda^2}{3!(4\pi)^4}\frac{ z x (1-z-x)[z(1-z)k_{T_c}^2+x(1-x)k_{T_d}^2]}{\Big[(z(1-z)k_{T_c}^2+x(1-x)k_{T_d}^2)^2-4 z^2 x^2k_{T_c}^2k_{T_d}^2\Big]^{3/2}},
\end{equation}
where we used $M_{n+c+d}=\frac{\lambda}{p_a^2}M_{n}$, and furthermore added $1/3!$ for the three identical particles.

\section{Cancelation of IR Divergences in \texorpdfstring{$\phi_{\text{4D}}^3$}{phi34d}}\label{sec:phi34dIRcanc}
We set $m=0$ and use dimensional regularization to regulate both the UV and the IR divergences~\cite{Gastmans:1973uv}. Let us find the total cross section of the process $\phi^*\to \phi\to X$, where $\phi$ is off-shell: 
\begin{align}\nonumber
\sigma_{total}=& R^2 \sigma_b+ \sigma_v + \sigma_r.
\end{align}

Here, $R$ is the field renormalization, $b$ stands for Born, $v$ for virtual, and $r$ for the real emission contributions. In the limit $m\to 0$, keeping $g$ fixed, IR divergences should cancel as shown in Fig.~\ref{fig:multiHIRcancel}.
  
 \begin{figure}[ht!]
   \centering
  \begin{eqnarray*}
\parbox{23mm}{\begin{fmffile}{IRdiv0}
    \begin{fmfgraph}(60, 30)
     \fmfleft{l1,l2,l3}
     \fmfright{r1,r2,r3}
     \fmf{dashes}{l2,ol1}
     \fmf{dashes}{ol1,or1}
     \fmf{dashes}{ol1,or3}
     \fmf{dashes}{or3,r3}
     \fmf{dashes}{or1,r1}
     \fmffreeze
     \fmf{dashes}{or1,or3}
    \end{fmfgraph}\end{fmffile}
  }
  \times\quad
  \parbox{23mm}{\begin{fmffile}{IRdiv0_1v2}
    \begin{fmfgraph}(60, 30)
     \fmfleft{l1,l2,l3}
     \fmfright{r1,r2,r3}
     \fmf{dashes}{l2,o}
     \fmf{dashes}{o,r3}
     \fmf{dashes}{o,r1}
    \end{fmfgraph}\end{fmffile}
  } ^* \quad \longleftrightarrow \qquad
  \parbox{23mm}{\begin{fmffile}{IRdiv01R1}
    \begin{fmfgraph}(60, 30)
     \fmfleft{l1,l2,l3}
     \fmfright{r1,r2,r3,r4,r5}
     \fmf{dashes,tension=2}{l2,ol1}
     \fmf{dashes,tension=2.5}{ol1,or1}
     \fmf{dashes}{or1,r5}
      \fmf{dashes,tension=2.5}{ol1,or2}
     \fmf{dashes}{or2,r1}
     \fmffreeze
     \fmf{dashes}{or1,r4}
    \end{fmfgraph}\end{fmffile}
 } \times \quad \parbox{23mm}{\begin{fmffile}{IRdiv0_1R2}
    \begin{fmfgraph}(60, 30)
     \fmfleft{l1,l2,l3}
     \fmfright{r1,r2,r3,r4,r5}
     \fmf{dashes,tension=2}{l2,ol1}
     \fmf{dashes,tension=2.5}{ol1,or1}
     \fmf{dashes}{or1,r5}
      \fmf{dashes,tension=2.5}{ol1,or2}
     \fmf{dashes}{or2,r1}
     \fmffreeze
     \fmf{dashes}{or2,r2}
    \end{fmfgraph}\end{fmffile}
 } ^*
  \end{eqnarray*}
  \begin{eqnarray*}
  \parbox{25mm}{\begin{fmffile}{IRdiv0_2V}
    \begin{fmfgraph}(60, 30)
     \fmfleft{l1,l2,l3}
     \fmfright{r1,r2,r3}
     \fmf{dashes}{l2,ol1}
     \fmf{dashes}{ol1,or1}
     \fmf{dashes}{or1,or2}
     \fmf{dashes}{ol1,or3}
     \fmf{dashes}{or3,or4}
     \fmf{dashes}{or4,r3}
     \fmf{dashes}{or2,r1}
     \fmffreeze
     \fmf{dashes,left,tension=0}{or3,or4}
    \end{fmfgraph}\end{fmffile}
  }  \times\quad
  \parbox{25mm}{\begin{fmffile}{IRdiv0_2v2}
    \begin{fmfgraph}(60, 30)
     \fmfleft{l1,l2,l3}
     \fmfright{r1,r2,r3}
     \fmf{dashes}{l2,o}
     \fmf{dashes}{o,r3}
     \fmf{dashes}{o,r1}
    \end{fmfgraph}\end{fmffile}
  } ^* \quad \longleftrightarrow \qquad \Bigg|\quad
  \parbox{25mm}{\begin{fmffile}{IRdiv0_2R}
    \begin{fmfgraph}(60, 30)
     \fmfleft{l1,l2,l3}
     \fmfright{r1,r2,r3,r4,r5}
     \fmf{dashes,tension=2}{l2,ol1}
     \fmf{dashes,tension=2.5}{ol1,or1}
     \fmf{dashes}{or1,r5}
      \fmf{dashes,tension=2.5}{ol1,or2}
     \fmf{dashes}{or2,r1}
     \fmffreeze
     \fmf{dashes}{or1,r4}
    \end{fmfgraph}\end{fmffile}
 } \Bigg|^2
      \end{eqnarray*}
      \caption{Cancelation of IR divergences between virtual (left) and real (right) contributions to the total cross section.}
      \label{fig:multiHIRcancel}
  \end{figure}

 Since the final particles are identical, we need to divide the cross sections by $n!$, where $n$ is the number of the final particles. We have
\begin{align}\nonumber
\sigma_b &= \frac{1}{2!} \int \Pi_2 |M_0|^2,\\\nonumber
\sigma_v &= \frac{1}{2!} \int \Pi_2 |\delta M_v|^2,\\\nonumber
\sigma_r &= \frac{1}{3!} \int \Pi_3 |M_R|^2.
\end{align}

The divergent part of the field strength tensor, $R$, is proportional to
\begin{align}\nonumber
 \int \frac{d^4k}{k^4}=\frac{1}{\epsilon}-\frac{1}{\epsilon'}.
\end{align} 

The $\epsilon$ is due to IR divergence and $\epsilon'$ due to UV divergence. Later on, both will cancel separately in the total cross section. But, for convenience, instead of keeping both UV and IR regulators, we can instead set $\epsilon=\epsilon'$~\cite{Schwartz:2013pla}, which gives $R=1$. 

For the born cross section we find
\begin{align}\nonumber
\sigma_b&= \frac{g^2}{16\pi}.
\end{align}

There are two virtual diagrams that, once interfere with the tree lever, give the next order correction to the total cross section. The first one is the vertex correction shown in Fig.~\ref{fig:multiHvertexcorrection}.
 \begin{figure}[ht!]
   \centering
{\begin{fmffile}{hhh_Loop0}
    \begin{fmfgraph*}(130, 65)
     \fmfleft{l1,l2,l3}
     \fmfright{r1,r2,r3}
     \fmf{dashes,label=$Q$}{l2,ol1}
     \fmf{dashes}{ol1,or1}
     \fmf{dashes}{ol1,or3}
     \fmf{dashes,label=$1$}{or3,r3}
     \fmf{dashes,label=$2$}{or1,r1}
     \fmffreeze
     \fmf{dashes}{or1,or3}
    \end{fmfgraph*}\end{fmffile}
  }
    \caption{Vertex correction.}
    \label{fig:multiHvertexcorrection}
  \end{figure}
  
 The amplitude for this vertex is
\begin{align}\nonumber
M_{v1}=\frac{-ig^3 }{(4\pi)^{2}}\frac{\Gamma(3-d/2)}{Q^2}(\frac{4\pi}{Q^2})^{2-d/2}\int dzdy \frac{1}{(-y z)^{3-d/2}},
\end{align}
where $d=4-\epsilon$, but $\epsilon<0$ so that the integral converges. Hence, the leading correction to the cross section is
\begin{align}\nonumber
\delta M^2=& M_0 M_{v1}^* +M_0^*M_{v1},\\\nonumber
=& +\frac{g^4 }{(4\pi)^{2}}\frac{2\Gamma(3-d/2)}{Q^2}(\frac{4\pi}{Q^2})^{2-d/2}\int dzdy \frac{1}{(-y z)^{3-d/2}}.
\end{align}
The phase-space integral in $d$ dimension is
\begin{align}\nonumber
\int \Pi_2 =(\frac{4\pi}{Q^2})^{2-d/2}\frac{1}{2^d\sqrt{\pi}\Gamma(\frac{d-1}{2})}.
\end{align}

We find
\begin{align}\nonumber
\sigma_{v1} =& \frac{1}{2} \int \Pi |\delta M_{v1}|^2,\\\nonumber
=& \frac{g^4 }{128 \pi^3Q^2}(\frac{4\pi e^{-\gamma_E}}{Q^2})^{4-d}\Big(-\frac{4}{\epsilon^2}-\frac{4}{\epsilon}-4+\frac{5\pi^2}{6}\Big),
\end{align}
where $\gamma_E$ is the Euler-Mascheroni constant. 

 \begin{figure}[ht!]
   \begin{center}
  \centering
\begin{fmffile}{hhh_Loop20}
    \begin{fmfgraph*}(150, 65)
     \fmfleft{l1,l2,l3}
     \fmfright{r1,r2,r3}
     \fmf{dashes,tension=2,label=$Q$}{l2,ol}
     \fmf{phantom,tension=2}{ol,o1}
     \fmf{dashes,tension=2}{o1,or}
     \fmf{dashes,label=$1$}{or,r3}
     \fmf{dashes,label=$2$}{or,r1}
     \fmffreeze
     \fmf{dashes,right,tension=0}{ol,o1}
     \fmf{dashes,left,tension=0}{ol,o1}
    \end{fmfgraph*}\end{fmffile}

    \caption{Vacuum correction.}
    \label{fig:multiHvacuumcorrection}
      \end{center}
  \end{figure}

The other loop diagram is the vacuum diagram for the intermediate Higgs shown in Fig.~\ref{fig:multiHvacuumcorrection}. The integral for this diagram is UV divergent but not IR divergent. The amplitude is
\begin{align}\nonumber
M_{v2}=&\frac{ig^3}{2(4\pi)^2Q^2}\Big(\frac{4\pi}{Q^2}\Big)^{2-d/2}\Gamma(2-d/2)\int \frac{dx}{(x(x-1))^{2-d/2}},
\end{align}
where the factor of two in the first denominator is the symmetry factor. We find
\begin{align}\nonumber
\sigma_{v2}=&\frac{g^4}{128\pi^3Q^2}\Big(\frac{4\pi e^{-\gamma_E}}{Q^2}\Big)^{4-d}\Big(-\frac{1}{\epsilon}-2\Big).
\end{align}

Altogether, the cross section is
\begin{align}\nonumber
\sigma_v=\sigma_{v1}+\sigma_{v2}=A_b\frac{g^4}{128\pi^3Q^2}\Big(\frac{4\pi e^{-\gamma_E}}{Q^2}\Big)^{4-d}\Big(-\frac{4}{\epsilon^2}-\frac{5}{\epsilon}-6+\frac{5\pi^2}{6}\Big).
\end{align}

 \begin{figure}[ht!]
   \centering
  \begin{eqnarray*}
\parbox{15mm}{\begin{fmffile}{hhh_Realemission0}
    \begin{fmfgraph*}(80, 85)
     \fmfleft{l1,l2,l3}
     \fmfright{r1,r2,r3,r4,r5}
     \fmf{dashes,label=$Q$}{l2,ol1}
     \fmf{dashes,tension=2.5}{ol1,or1}
     \fmf{dashes,label=$1$}{or1,r4}
     \fmf{dashes,label=$2$}{or1,r2}
     \fmffreeze
     \fmf{phantom}{or2,r5}
     \fmf{dashes,label=$3$,label.side=left}{ol1,or2}
    \end{fmfgraph*}\end{fmffile}
  }\hspace{2cm}+\quad\parbox{20mm}{\begin{fmffile}{hhh_Realemission10}
    \begin{fmfgraph*}(80, 85)
     \fmfleft{l1,l2,l3}
     \fmfright{r1,r2,r3,r4,r5}
     \fmf{dashes,tension=2.5,label=$Q$}{l2,ol1}
     \fmf{dashes,tension=2.5}{ol1,or1}
     \fmf{dashes,label=$1$,label.side=left}{or1,r4}
      \fmf{dashes,tension=2.5}{ol1,or2}
     \fmf{dashes,label=$2$,label.side=left}{or2,r2}
     \fmffreeze
     \fmf{dashes,label=$3$,label.side=right}{or2,r1}
    \end{fmfgraph*}\end{fmffile}
 }
    \hspace{1.5cm}+\quad\parbox{20mm}{\begin{fmffile}{hhh_Realemission20}
    \begin{fmfgraph*}(80, 85)
     \fmfleft{l1,l2,l3}
     \fmfright{r1,r2,r3,r4,r5}
     \fmf{dashes,tension=2.5,label=$Q$}{l2,ol1}
     \fmf{dashes,tension=2.5}{ol1,or1}
     \fmf{dashes,label=$1$,label.side=left}{or1,r4}
      \fmf{dashes,tension=2.5}{ol1,or2}
     \fmf{dashes,label=$2$,label.side=right}{or2,r2}
     \fmffreeze
     \fmf{dashes,label=$3$,label.side=right}{or1,r5}
    \end{fmfgraph*}\end{fmffile}
  }
    \end{eqnarray*}
    \caption{The real emission.}
    \label{fig:multiHrealemission}
  \end{figure}

For the real emission, we have three diagrams, shown in Fig.~\ref{fig:multiHrealemission}. The amplitude is given by
\begin{align}\nonumber
M_r=& \frac{-ig^2}{Q^2}\Bigg(\frac{-1+x_1x_2+(2-x_1-x_2)(x_1+x_2)}{(1-x_1)(1-x_2)(x_1+x_2-1)}\Bigg),
\end{align}

where $x_i=\frac{2E_i}{Q}$, and $x_1+x_2+x_3=2$. The 3-body phase-space integral in dimensional regularization is

\begin{align}\nonumber
\int d\Pi_3 =&\Big(\frac{Q^2}{4\pi}\Big)^{d-4}\frac{Q^2}{128\pi^3 \Gamma(d-2)}\times\nonumber\\\nonumber
&\int_0^1dx_1\int_{1-x_1}^1dx_2\frac{1}{((1-x_1)(1-x_2)(x_1+x_2-1))^{2-d/2}}.
\end{align}
A nice way of doing the integral is by changing the variable $x_2=1-y x_1$, $0<y<1$. We find
\begin{align}\nonumber
\sigma_r=&\frac{1}{3!} \int d\Pi_3 |M_R|^2,\\\nonumber
=& \frac{g^4}{128\pi^3 Q^2}\Big(\frac{4\pi e^{-\gamma_E}}{Q^2}\Big)^{4-d}\Bigg( \frac{4}{\epsilon^2}+\frac{5}{\epsilon}+\frac{9}{2}-\frac{5\pi^2}{6}\Bigg).
\end{align}

The sum of the to two contributions to the total cross sections is
\begin{align}\nonumber
\sigma_v+\sigma_r =& \frac{g^4}{128\pi^3 Q^2}\Big(\frac{4\pi e^{-\gamma_E}}{Q^2}\Big)^{4-d}\Big( -\frac{3}{2} \Big).
\end{align}

Hence, after setting $d=4$, 
\begin{align}\nonumber
\sigma_{\text{total}}= \frac{g^2}{16\pi}\Bigg(1-\frac{3g^2}{16\pi^2 Q^2}\Bigg),
\end{align}

which is a finite number.

\section{AP Function for \texorpdfstring{$\phi^3_{\text{4D}}$}{phi34dapp} }\label{atphi34d}

We start by writing the cross section for production of $n+1$ particles in terms of the cross section of production of $n$ particles plus a split particles, labeled $c$ (Fig.~\ref{fdiag:phi3split}), that is radiated from one of the final legs. We have

 \begin{figure}[ht!]
   \centering
\begin{fmffile}{h_to_hh0}
    \begin{fmfgraph*}(150, 65)
     \fmfleft{l}
     \fmfright{r1,r2}
     \fmf{dashes,tension=2.5,label=$a$}{l,o}
     \fmf{dashes,label=$b$}{o,r2}
     \fmf{dashes,label=$c$}{o,r1}
    \end{fmfgraph*}\end{fmffile}

  \caption{$\phi\to \phi\phi$. Momentums are flowing to the right. ``c" is integrated over. }
  \label{fdiag:phi3split}
  \end{figure}

\begin{align}\nonumber
\sigma_{1\to n+c}&=\text{flux factor} \int d \Pi^f d \Pi^c \quad | M_{n+c}|^2, \\\nonumber
& = \text{flux factor} \int d \Pi^f dz d k_T^2 \mathcal{P}(z,k_T) |M_n|^2, \\
& = \sigma_{1\to n} \int dz dk_T^2 \mathcal{P}_{\phi^3}(z,k_T),\label{eq:APdefinition}
\end{align}
where $c$ is assumed to be collinear to the leg labeled $b$. We have defined $z=E_b/E_c$. The $\PP$ is the splitting function. 

We write the momentum as (called Sudakov decomposition~\cite{Ellis:1991qj}),
\begin{align}\nonumber
p_b &=zp_a+\beta u + k_T,\\\nonumber
p_c &= (1-z)p_a -\beta u -k_T,
\end{align}
so that $p_a=p_b+p_c$. The vector $u$ is an arbitrary vector perpendicular to $k_T$, we choose it to be $(1,0,0,-1)$. The phase-space integral for particle $c$ becomes
\begin{align}\nonumber
d\Pi^c &=\frac{dz k_T d k_T d\phi d\beta}{(2\pi)^4} \times J (=E_a+p_{a3})\times (2\pi)\delta (p_c^2-m^2),\\\nonumber
&=\frac{dz d k_T^2}{4(2\pi)^2 (1-z)},
\end{align}
where in the second line we have integrated over $\beta$ and $\phi$. We further have
\begin{equation}\label{eq:pa2phi3}
p_a^2=p_b^2+p_c^2+2p_bp_c=\frac{k_t^2}{z(1-z)}+\frac{p_b^2}{(1-z)}+\frac{p_c^2}{z},\qquad k_T^2,p_{b,c}^2 \ll E_{b,c}. 
\end{equation}

Hence,
\begin{equation}\nonumber
\PP_{\phi^3} = \frac{g^2}{4(2\pi)^2}\frac{1}{z(1-z)}\frac{1}{(p_a^2-m^2)^2}.
\end{equation}
The extra $1/z$ in the first fraction comes from changing the phase-space factor of particle $b$ to that of particle $a$ (Fig.~\ref{fdiag:phi3split}). Using \eqref{eq:pa2phi3} we find
\begin{align}\nonumber
\PP_{\phi^3} &=\frac{g^2}{16\pi^2}\frac{z(1-z)}{\Big(k_T^2+z p_b^2+(1-z) p_c^2 -z(1-z)m^2 \Big)^2},\\
&=\frac{g^2}{16\pi^2}\frac{z(1-z)}{\Big(k_T^2+ m^2 (1-z+z^2)\Big)^2}.
\end{align}

\section{Comparison to Fix-Order Calculation}\label{sec:fixorder}

\begin{figure}[ht!]
\begin{center}
\includegraphics[width=.7\textwidth]{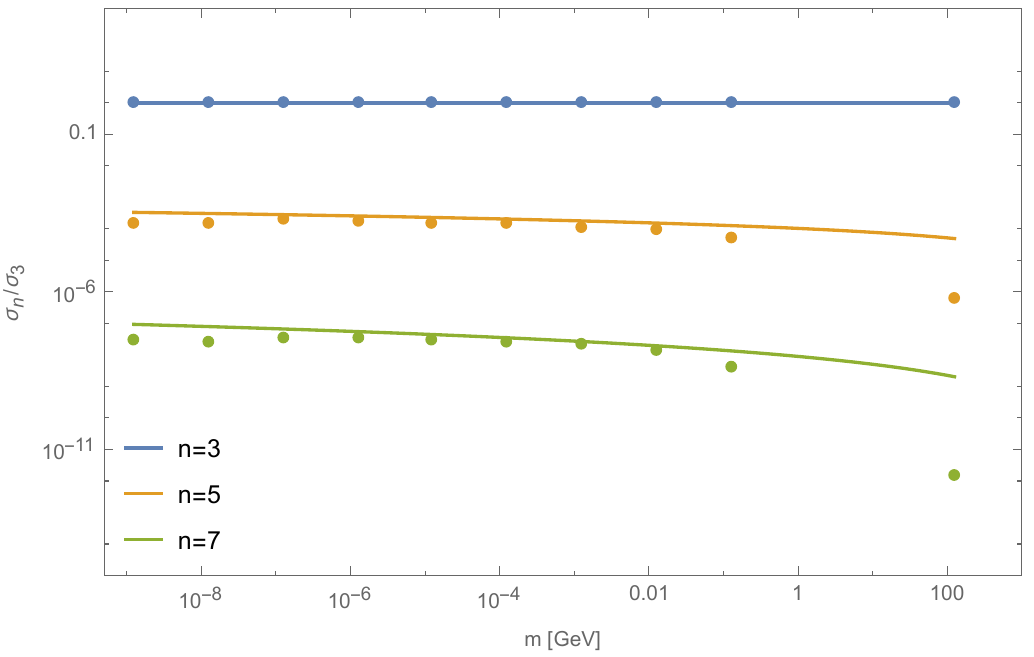}
\caption{Madgraph computation vs. Jet calculation of $gg\to \phi^*\to n\phi$ in the $\phi^4$ theory. From top to bottom: n = 3,5, and 7.}
\label{fig:phi4madgraph}
\end{center}
\end{figure}

As argued, we can use the mass as the resolution parameter and turn the jet rates into particle cross sections \eqref{eq:ratephi36dexpandoffshell}~-~\eqref{eq:ratephi34dexpandoffshell}. In these expressions the total cross section in unknown, hence, we find the ratios $\sigma_n/\sigma_2$ and $\sigma_n/\sigma_3$ for $\phi^3$ and $\phi^4$ theories respectively. We can then utilities MADGRAPH~\cite{Alwall:2014hca} to compute the cross section for the process $gg\to \phi^*\to$few $ \phi$. We used the model $HEFT$ and turned off (by simply erasing the interaction in \emph{vertices.py} file) cubic interactions for analyzing $\phi^4$ theory and, vice versa, turned off the quartic coupling to study $\phi^3$ theory.

\begin{table}[h!]
\begin{center}
\begin{tabular}{c|l|| r|l||r|l}
\hline
\multicolumn{2}{c||}{$\phi^3$}&\multicolumn{2}{c||}{$\phi^4$}&\multicolumn{2}{c}{Broken Theory}\\
\hline
n &\# of diagrams&n&\# of diagrams&n&\# of diagrams\\
\hline
2&1&3&1&2&1\\
3&3& 5& 10&3&4\\
4&15& 7& 280&4&25\\
5&105&9& 15400&5&220\\
6&945& & &6& 2485\\
7&10395& & &&
\end{tabular}
\caption{Number of Feynman diagrams for the process $gg\to\phi^*\to n\phi$ for different theories.}
\label{Tab1}
\end{center}
\end{table}

Due to the huge number of Feynman diagram, listed in Table.~\ref{Tab1}, it is not possible\footnote{In a reasonable amount of time using a home computer.} to compute more than few particles in the final state. 
\begin{figure}[ht!]
\centering
\begin{subfigure}[l]{0.48\textwidth}	
\includegraphics[width=.99\textwidth]{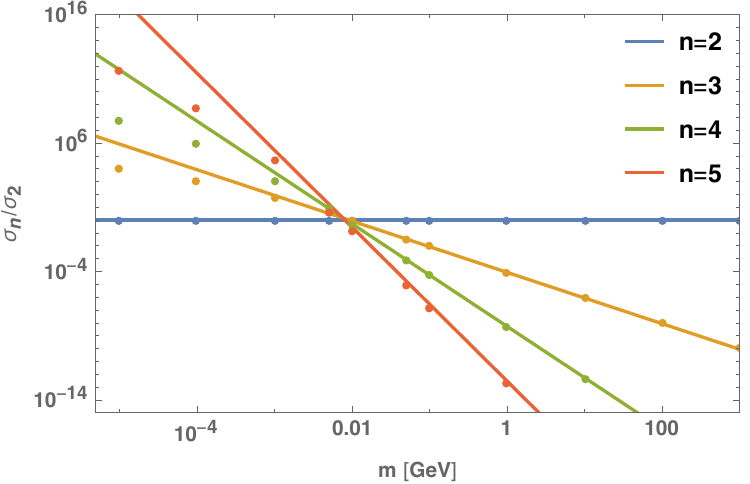}
\caption{$g=1/8$}
\label{fig:phi3madgraph-1}
\end{subfigure}
\begin{subfigure}[r]{0.48\textwidth}	
\includegraphics[width=.99\textwidth]{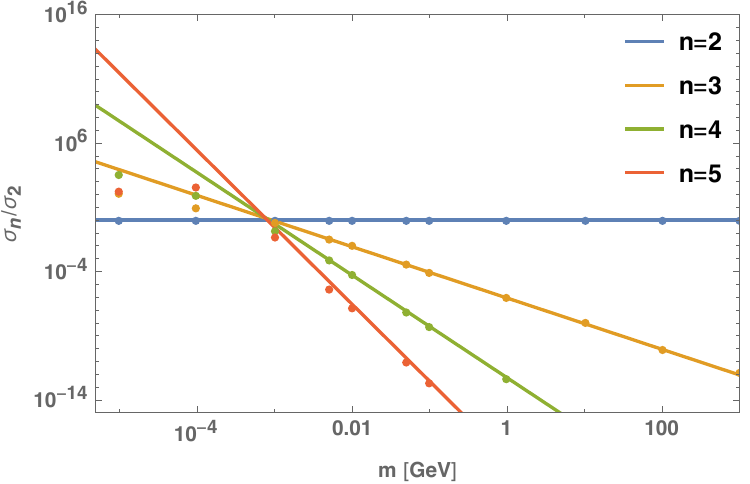}
\caption{$g=1/80$}
\label{fig:phi3madgraph-2}
\end{subfigure}
\caption{Madgraph computation vs. Jet calculation of $gg\to \phi^*\to n\phi$ in the $\phi^3$ theory.}
\label{fig:phi3madgraph}
\end{figure}

In Fig.~\ref{fig:phi4madgraph}, we have computed the cross section as a function of the scalar mass for $\phi^4$ theory \eqref{eq:ratephi44dexpandoffshell}, while fixing the center-of-mass energy at $10$ TeV. We can clearly see that our approximation becomes better at smaller $m$.

\begin{figure}[ht!]
\centering
\includegraphics[width=.7\textwidth]{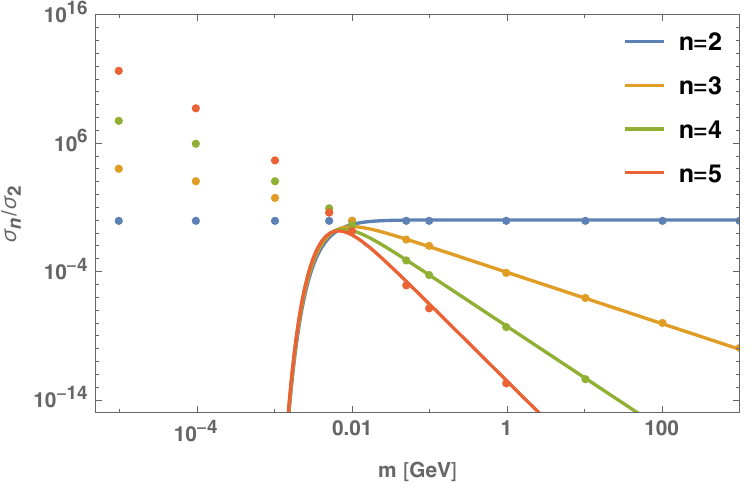}
\caption{Madgraph computation vs. Jet calculation of $gg\to \phi^*\to n\phi$ in the $\phi^3$ theory with the inclusion of Sudakov factors. $g=1/8$.}
\label{fig:phi3madgraph-3}
\end{figure}

In Fig.~\ref{fig:phi3madgraph}, we have repeated the calculation for the $\phi^3$ theory, \eqref{eq:ratephi34dexpandoffshell}. We again see that the approximation becomes better at smaller mass. However, here we see that for $g=1/8$, at around $m=0.01$ GeV, the predication fails. At this point the ratio $g^2/(4\pi)^2m^2$ becomes larger than 1, and we cannot trust the expansion that leads to \eqref{eq:ratephi34dexpandoffshell}. In Fig.~\ref{fig:phi3madgraph-2} we have shown that this is the case for $g=1/80$ as well. If we don't expand the Sudakov factors and use \eqref{eq:Rnph36d2-1}, all the cross sections will decrease rapidly at these points, as shown in Fig.~\ref{fig:phi3madgraph-3}.

\end{appendices}



\newpage


\nocite{*}
\begin{thebibliography}{99}
%
\bibitem{Ringwald:1989ee}
A.~Ringwald,
``High-Energy Breakdown of Perturbation Theory in the Electroweak Instanton Sector,''
Nucl. Phys. B \textbf{330}, 1-18 (1990)
%
\bibitem{Espinosa:1989qn}
O.~Espinosa,
``High-Energy Behavior of Baryon and Lepton Number Violating Scattering Amplitudes and Breakdown of Unitarity in the Standard Model,''
Nucl. Phys. B \textbf{343}, 310-340 (1990)
 %
  \bibitem{zakharov} 
 V.I. Zakharov,
``Unitary constraints on multiparticle weak production,"
\href{http://www.sciencedirect.com/science/article/pii/055032139190322O}{Nuclear Physics B,
Volume 353, Issue 3,
1991,}
%
%
\bibitem{Argyres:1992np}
E.~N.~Argyres, R.~H.~P.~Kleiss and C.~G.~Papadopoulos,
``Amplitude estimates for multi - Higgs production at high-energies,''
Nucl. Phys. B \textbf{391}, 42-56 (1993)
%
\bibitem{Brown:1992ay} 
 L.~S.~Brown,
 ``Summing tree graphs at threshold,''
 Phys.\ Rev.\ D {\bf 46}, R4125 (1992)
 [\arXivold{hep-ph/9209203}].
%
\bibitem{Smith:1992rq}
B.~H.~Smith,
``Summing one loop graphs in a theory with broken symmetry,''
Phys. Rev. D \textbf{47}, 3518-3520 (1993)
[\arXiv{hep-ph/9209287}{hep-ph}].
%
\bibitem{Voloshin:1994yp} 
 M.~B.~Voloshin,
 ``Nonperturbative methods,''
 [\arXivold{hep-ph/9409344}].
 %
 \bibitem{Libanov:1994ug} 
 M.~V.~Libanov, V.~A.~Rubakov, D.~T.~Son and S.~V.~Troitsky,
 ``Exponentiation of multiparticle amplitudes in scalar theories,''
 Phys.\ Rev.\ D {\bf 50}, 7553 (1994)
 [\arXivold{hep-ph/9407381}].
%
\bibitem{Libanov:1997nt}
M.~V.~Libanov, V.~A.~Rubakov and S.~V.~Troitsky,
``Multiparticle processes and semiclassical analysis in bosonic field theories,''
Phys. Part. Nucl. \textbf{28}, 217-240 (1997)
%
\bibitem{Schenk:2021yea}
S.~Schenk,
``The breakdown of resummed perturbation theory at high energies,''
JHEP \textbf{03}, 100 (2022)
[\arXiv{2109.00549}{hep-ph}].
%
%
\bibitem{Khoze:2022fbf}
V.~V.~Khoze and S.~Schenk,
``Multiparticle amplitudes in a scalar EFT,''
JHEP \textbf{05}, 134 (2022)
[\arXiv{2203.03654}{hep-th}].
%
\bibitem{Argyres:1993wz} 
 E.~N.~Argyres, R.~H.~P.~Kleiss and C.~G.~Papadopoulos,
 ``Multiscalar amplitudes to all orders in perturbation theory,''
 Phys.\ Lett.\ B {\bf 308}, 292 (1993)
 Addendum: [Phys.\ Lett.\ B {\bf 319}, 544 (1993)]
 [\arXivold{hep-ph/9303321}].
%
%
\bibitem{Papadopoulos:1993aw}
C.~G.~Papadopoulos, R.~H.~P.~Kleiss and E.~N.~Argyres,
``Multiscalar amplitudes in perturbation theory,''
%

\bibitem{Son:1995wz}
D.~T.~Son,
``Semiclassical approach for multiparticle production in scalar theories,''
Nucl. Phys. B \textbf{477}, 378-406 (1996)
[\arXiv{hep-ph/9505338}{hep-ph}].
%
%
\bibitem{Khoze:2018mey}
V.~V.~Khoze and J.~Reiness,
``Review of the semiclassical formalism for multiparticle production at high energies,''
Phys. Rept. \textbf{822}, 1-52 (2019)
[\arXiv{1810.01722}{hep-ph}].
%
\bibitem{Demidov:2022ljh}
S.~V.~Demidov, B.~R.~Farkhtdinov and D.~G.~Levkov,
``Suppression exponent for multiparticle production in \ensuremath{\lambda}\ensuremath{\phi}$^{4}$ theory,''
JHEP \textbf{02} (2023), 205
[\arXiv{2212.03268}{hep-ph}].
%
\bibitem{Lipatov:1977hj}
L.~N.~Lipatov,
``Divergence of Perturbation Series and Pseudoparticles,''
JETP Lett. \textbf{25}, 104-107 (1977)
%
 \bibitem{maggshif91}
Michele Maggiore, Mikhail Shifman,
``Where have all form factors gone in the instanton amplitudes?,"
\href{http://www.sciencedirect.com/science/article/pii/055032139190610A}{Nuclear Physics B,
Volume 365, Issue 1,
1991,}
%
 \bibitem{maggshif92}
 Michele Maggiore, Mikhail Shifman,
``Multiparticle production in weak coupling theories in the Lipatov approach,"
\href{http://www.sciencedirect.com/science/article/pii/055032139290513B}{Nuclear Physics B,
Volume 380, Issues 1–2,
1992,}

%
 \bibitem{cornwall92}
John M. Cornwall, George Tiktopoulos,
``Functional Schrödinger equation approach to high-energy multiparticle scattering,"
\href{http://www.sciencedirect.com/science/article/pii/037026939290501T}{Physics Letters B,
Volume 282, Issues 1–2,
1992,}
%
\bibitem{Cornwall:1993rh} 
 J.~M.~Cornwall and G.~Tiktopoulos,
 ``Semiclassical matrix elements for the quartic oscillator,''
 Annals Phys.\ {\bf 228}, 365 (1993).
%
\bibitem{Jaeckel:2018ipq} 
 J.~Jaeckel and S.~Schenk,
 ``Exploring High Multiplicity Amplitudes in Quantum Mechanics,''
 Phys.\ Rev.\ D {\bf 98}, no. 9, 096007 (2018)
 [\arXiv{1806.01857}{hep-ph}].
%
\bibitem{Jaeckel:2018tdj}
J.~Jaeckel and S.~Schenk,
``Exploring high multiplicity amplitudes: The quantum mechanics analogue of the spontaneously broken case,''
Phys. Rev. D \textbf{99}, no.5, 056010 (2019)
doi:10.1103/PhysRevD.99.056010
[\arXiv{1811.12116}{hep-ph}].
%
\bibitem{Khoze:2014kka} 
 V.~V.~Khoze,
 ``Perturbative growth of high-multiplicity W, Z and Higgs production processes at high energies,''
 JHEP {\bf 1503}, 038 (2015)
 [\arXiv{1411.2925}{hep-ph}].
%
\bibitem{Khoze:2015yba}
V.~V.~Khoze,
``Diagrammatic computation of multi-Higgs processes at very high energies: Scaling log $?_n$ with MadGraph,''
Phys. Rev. D \textbf{92}, no.1, 014021 (2015)
[\arXiv{1504.05023}{hep-ph}].
%
\bibitem{Degrande:2016oan} 
 C.~Degrande, V.~V.~Khoze and O.~Mattelaer,
 ``Multi-Higgs production in gluon fusion at 100 TeV,''
 Phys.\ Rev.\ D {\bf 94}, 085031 (2016)
 [\arXiv{1605.06372}{hep-ph}].
%
\bibitem{Khoze:2017tjt} 
 V.~V.~Khoze and M.~Spannowsky,
 ``Higgsplosion: Solving the Hierarchy Problem via rapid decays of heavy states into multiple Higgs bosons,''
 Nucl.\ Phys.\ B {\bf 926}, 95 (2018)
 [\arXiv{1704.03447}{hep-ph}].
%
\bibitem{Khoze:2017ifq} 
 V.~V.~Khoze,
 ``Multiparticle production in the large $\lambda n$ limit: realising Higgsplosion in a scalar QFT,''
 JHEP {\bf 1706}, 148 (2017)
 [\arXiv{1705.04365}{hep-ph}].
%
\bibitem{Voloshin:2017flq} 
 M.~B.~Voloshin,
 ``Loops with heavy particles in production amplitudes for multiple Higgs bosons,''
 Phys.\ Rev.\ D {\bf 95}, no. 11, 113003 (2017)
 [\arXiv{1704.07320}{hep-ph}].
 %
 %
\bibitem{Belyaev:2018mtd}
A.~Belyaev, F.~Bezrukov, C.~Shepherd and D.~Ross,
``Problems with Higgsplosion,''
Phys. Rev. D \textbf{98}, no.11, 113001 (2018)
[\arXiv{1808.05641}{hep-ph}].
%
\bibitem{Monin:2018cbi}
A.~Monin,
``Inconsistencies of higgsplosion,''
[\arXiv{1808.05810}{hep-th}].
%
\bibitem{Khoze:2018qhz}
V.~V.~Khoze and M.~Spannowsky,
``Consistency of Higgsplosion in Localizable QFT,''
Phys. Lett. B \textbf{790}, 466-474 (2019)
[\arXiv{1809.11141}{hep-ph}].
%
 \bibitem{Dine:2020ybn}
M.~Dine, H.~H.~Patel and J.~F.~Ulbricht,
``Behavior of Cross Sections for Large Numbers of Particles,''
[\arXiv{2002.12449}{hep-ph}].
%
\bibitem{Curko:2019dtu}
A.~Curko and G.~Cynolter,
``Unitarity in Multi-Higgs Production,''
[\arXiv{1911.04784}{hep-ph}].
%
\bibitem{Dokshitzer:1991wu}
Y.~L.~Dokshitzer, V.~A.~Khoze, A.~H.~Mueller and S.~I.~Troian,
``Basics of perturbative QCD,''
%
%
\bibitem{Ellis:1991qj} 
 R.~K.~Ellis, W.~J.~Stirling and B.~R.~Webber,
 ``QCD and collider physics,''
 Camb.\ Monogr.\ Part.\ Phys.\ Nucl.\ Phys.\ Cosmol.\ {\bf 8}, 1 (1996).
%
%
\bibitem{Altarelli:1977zs} 
 G.~Altarelli and G.~Parisi,
 ``Asymptotic Freedom in Parton Language,''
 Nucl.\ Phys.\ B {\bf 126}, 298 (1977).
%
\bibitem{Lipatov:1974qm} 
 L.~N.~Lipatov,
 ``The parton model and perturbation theory,''
 Sov.\ J.\ Nucl.\ Phys.\ {\bf 20}, 94 (1975);
%
%
\bibitem{Sudakov:1954sw} 
 V.~V.~ Sudakov,
 ``Vertex parts at very high-energies in quantum electrodynamics,''
 Sov.\ Phys.\ JETP {\bf 3}, 65 (1956)
 %
 \bibitem{collins:1989} 
  J.~C.~Collins,
 `` Sudakov form-factors,''
 Adv.\ Ser.\ Direct.\ High Energy Phys.\ {\bf 5}, 573 (1989)
 [\arXivold{hep-ph/0312336}].
 %
\bibitem{chang:1977}
S.~J. Chang and Y. P. Yao,
Nonperturbative approach to infrared behavior for ${({\ensuremath{\varphi}}^{3})}_{6}$ theory and a mechanism of confinement,
  Phys.\ Rev.\ D.16.2948 (1977).
%
\bibitem{Taylor:1978hu}
J.~C.~Taylor,
``Structure of High-energy Jets in Six-dimensional $\phi^3$ Theory,''
Phys. Lett. B \textbf{73}, 85-86 (1978)
%
\bibitem{Konishi:1979cb}
K.~Konishi, A.~Ukawa and G.~Veneziano,
``Jet Calculus: A Simple Algorithm for Resolving QCD Jets,''
Nucl. Phys. B \textbf{157}, 45-107 (1979)
%
\bibitem{Catani:1991hj}
S.~Catani, Y.~L.~Dokshitzer, M.~Olsson, G.~Turnock and B.~R.~Webber,
``New clustering algorithm for multi - jet cross sections in e+ e- annihilation,''
Phys. Lett. B \textbf{269}, 432-438 (1991)
%
 \bibitem{Gribov:1972}
 V.~N.~Gribov and L.~N.~Lipatov,
 ``Deep inelastic e p scattering in perturbation theory,''
 Sov.\ J.\ Nucl.\ Phys.\ {\bf 15}, 438 (1972);
 %

 \bibitem{Dokshitzer:1977}
 Y.~L.~Dokshitzer,
 ``Calculation of the Structure Functions for Deep Inelastic Scattering and e+ e- Annihilation by Perturbation Theory in Quantum Chromodynamics.,''
 Sov.\ Phys.\ JETP {\bf 46}, 641 (1977)
%
\bibitem{Plehn:2015dqa}
 T.~Plehn,
 ``Lectures on LHC Physics,''
 Lect.\ Notes Phys.\ {\bf 844}, 1 (2012)
 [\arXiv{0910.4182}{hep-ph}].
%
\bibitem{Bloch:1937pw} 
 F.~Bloch and A.~Nordsieck,
 ``Note on the Radiation Field of the electron,''
 Phys.\ Rev.\ {\bf 52}, 54 (1937).
%
\bibitem{Kinoshita:1962ur} 
 T.~Kinoshita,
 ``Mass singularities of Feynman amplitudes,''
 J.\ Math.\ Phys.\ {\bf 3}, 650 (1962);
 T.~D.~Lee and M.~Nauenberg,
 ``Degenerate Systems and Mass Singularities,''
 Phys.\ Rev.\ {\bf 133}, B1549 (1964).
%
\bibitem{Peskin:1995ev}
M.~E.~Peskin and D.~V.~Schroeder,
\emph{An Introduction to quantum field theory}, Addison-Wesley, 1995,
%
\bibitem{Schwartz:2013pla} 
 M.~D.~Schwartz,
 \emph{Quantum Field Theory and the Standard Model}, Cambridge University Press, 2014.
%
\bibitem{Srednicki:2007qs} 
 M.~Srednicki,
 \emph{Quantum field theory}, Cambridge University Press, 2007.
 %
\bibitem{Becher:2014oda}
T.~Becher, A.~Broggio and A.~Ferroglia,
``Introduction to Soft-Collinear Effective Theory,''
[\arXiv{1410.1892}{hep-ph}].
%
\bibitem{Chen:2016wkt} 
 J.~Chen, T.~Han and B.~Tweedie,
 ``Electroweak Splitting Functions and High Energy Showering,''
 JHEP {\bf 1711}, 093 (2017)
 [\arXiv{1611.00788}{hep-ph}].
ett.\ B {\bf 269}, 432 (1991).
%
\bibitem{Gastmans:1973uv} 
 R.~Gastmans and R.~Meuldermans,
 ``Dimensional regularization of the infrared problem,''
 \href{https://www.sciencedirect.com/science/article/pii/0550321373901466}{Nucl.\ Phys.\ B {\bf 63}, 277 (1973).}
%
 \bibitem{Alwall:2014hca} 
 J.~Alwall {\it et al.},
 ``The automated computation of tree-level and next-to-leading order differential cross sections, and their matching to parton shower simulations,''
 JHEP {\bf 1407}, 079 (2014)
 [\arXiv{1405.0301}{hep-ph}].
%
\bibitem{Brown:1990nm}
N.~Brown and W.~J.~Stirling,
``Jet cross sections at leading double logarithm in e+ e- annihilation,''
Phys. Lett. B \textbf{252}, 657-662 (1990)
%
\bibitem{mathematica}
Wolfram Research, Inc., Mathematica, Version 14.0, Champaign, IL (2024).
%
\end{thebibliography}
\end{document}